\documentclass[epj]{svjour}
\usepackage{latexsym}
\usepackage{graphics}
\usepackage[T1]{fontenc}
\usepackage[latin1]{inputenc}
\usepackage{graphicx}
\usepackage[english]{babel}
\usepackage{graphicx}
\usepackage{bm}
\usepackage{amsmath}
\usepackage{amssymb}
\usepackage{amsfonts}
\usepackage{epsfig}
\usepackage{colordvi}
\usepackage{color}
\usepackage{longtable}
\begin{document}
%
\title{Hamiltonian dynamics and Noether symmetries in Extended Gravity Cosmology}

\author{S. Capozziello\inst{1,2}, M. De Laurentis\inst{1,2}, S.D. Odintsov\inst{3,4}}

\institute{Dipartimento di Scienze Fisiche, Universit\`a di Napoli "Federico II \and INFN sez. di Napoli Compl. Univ. di Monte S. Angelo, Edificio G, Via Cinthia, I-80126 - Napoli, Italy \and  
Istitucio Catalana de Recerca i Estudis Avancats (ICREA) and Consejo Superior de Investigaciones Cientificas, ICE/CSIC and Institut de Ciencies de l Espai (IEEC-CSIC),
Campus UAB, Facultat de Ciencies, Torre C5-Par-2a pl, E-08193 Bellaterra (Barcelona), Spain.}
\date{Received: date / Revised version: date}

\abstract{We discuss the Hamiltonian dynamics for cosmologies coming from  Extended Theories of Gravity. In particular, minisuperspace models   are taken into account searching for   Noether symmetries.   The existence of conserved quantities  gives  selection rule to recover classical behaviors in cosmic evolution according to the so called Hartle criterion, that allows  to select correlated regions in the configuration space of dynamical variables. We show that  
 such a statement works for general classes of  Extended Theories of Gravity and is  conformally preserved.  Furthermore, the presence of Noether symmetries allows a straightforward classification of singularities that represent the points where the symmetry is broken.
 Examples  for nonminimally coupled and higher-order models are discussed. 
} 

\authorrunning{ S. Capozziello et al.} 
\titlerunning{\it Hamiltonian Dynamics in Extended Gravity Cosmology}
\maketitle

\tableofcontents

\vspace{6.mm}

\noindent 
\footnotesize{$^4$Also at Tomsk State Pedagogical University, Tomsk, Russia, and Eurasian National University, Astana, Kazakhstan.}

\section{Introduction}
Different points of view can be assumed in order to deal with Quantum
Cosmology. It can be considered as the first step towards the
construction of a complete theory of Quantum Gravity. Moreover, its goal is 
to find out  the law of  initial conditions from which our {\it classical}
universe started its evolution. However, with respect to other theories of
physics as Electromagnetism, General Relativity (GR) or 
Quantum Mechanics, boundary conditions for the evolution of the
system {\it universe} cannot be obviously set from {\it outside}. In standard theories,  the approach is to search for
 some field equations  ({\it e.g.} Maxwell's or Einstein's
equations or  Schr\"odinger's equation) and then  {\it impose},
from the outside, the laws of initial or boundary  conditions (the Cauchy problem). In Cosmology, by
definition, there is no {\it rest of the universe} so that
boundary conditions must be  {\it fundamental laws of physics}. In
this sense, a part the fact that Quantum Cosmology is a viable
outline to achieve Quantum Gravity, it can be considered as an
autonomous branch of physics due to the problem of finding initial
conditions \cite{hartle1}.

However, not only the conceptual difficulties, but also
mathematical ones make Quantum Cosmology difficult to handle. For
example, the superspace of {\it geometrodynamics} \cite{wheeler} has
infinite degrees of freedom so that it is not possible
to fully integrate the Wheeler-De Witt (WDW) equation. Moreover,
the Hilbert space of states describing  {\it universes}  is not
available and then it is not clear how to interpret
the solutions of WDW equation in the framework of probability
theory \cite{deWitt} .

Despite of these  shortcomings, several 
results have been obtained and Quantum Cosmology has become a sort
of {\it paradigm} in theoretical physics. For example
the infinite-dimensional superspace can be restricted to
suitable  finite-dimensional configuration spaces, the so- called {\it
minisuperspaces}. In this case, the above mathematical difficulties
can be circumvented since
the WDW equation reduces to a partial differential  equation and, in principle,  can be integrated.
The initial value problem can be approached  in some simplified ways as, for example, 
 the so called {\it no boundary condition} by Harte and Hawking
\cite{HH} and the {\it tunneling from nothing} by Vilenkin
\cite{vilenkin}. Both schemes give reasonable laws for initial conditions from
which our {\it classical} universe could be started. However also other approaches are possible \cite{kiefer}.
 However, it is better to stress that Quantum Cosmology is not fully satisfactory in view of  solving Quantum Gravity issues but 
 is a useful working scheme despite of different interpretations of results.
 
For example, the {\it Hartle criterion} \cite{hartle2} is an interpretative scheme for
the solutions of the WDW equation. Hartle proposed to look for peaks
of the wave function of the universe: if it is strongly peaked,
we have correlations among the geometrical
and matter degrees of freedom; if it is not peaked, correlations
are lost. In the first case, the emergence of classical
trajectories ({\it i.e.} universes) is expected. The analogy to 
non-relativistic Quantum Mechanics is  straightforward.  If  we have a
 a wave function, solution of the
Schr\"odinger equation in presence of a potential barrier,  an oscillatory regime is possible on and
outside the barrier;  a decreasing exponential behavior is present
under the barrier. The situation is analogous in Quantum
Cosmology: now the potential barrier has to be replaced by the
superpotential $U(h^{ij}, \phi)$, where $h^{ij}$ are the components of the
$3$-metric of geometrodynamics and $\phi$ is a generic
scalar field describing the matter content. More precisely, the
wave function of the universe can be written as
 \begin{equation}\label{1}
 \Psi[h_{ij}(x), \phi(x)]\sim e^{im_P^2{\cal S}}\,{,}
 \end{equation}
where $m_P$ is the Planck mass and
 \begin{equation}\label{2}
 {\cal S}\equiv {\cal S}_0+m_P^{-2}{\cal S}_1+O(m_P^{-4})\,{,}
 \end{equation}
 is the action which can be expanded.
We have to note that there is no normalization factor due to the lack
of a probability interpretative scheme.
 Inserting ${\cal S}$ into the WDW
equation (thta we will derive below) and equating similar powers of $m_P$, one obtains the
Hamilton-Jacobi equation for ${\cal S}_0$. Similarly, one gets equations
for ${\cal S}_1, {\cal S}_2,\ldots$, which can be solved considering results of
previous orders. We need only ${\cal S}_0$ to recover the
semi-classical limit of Quantum Cosmology \cite{halliwell}. If
$S_0$ is a real number, we get oscillating WKB modes and the
Hartle criterion is recovered since $\Psi$ is peaked on a
phase-space region defined by
 \begin{equation}\label{3}
 \pi_{ij}=m_P^2\,\frac{\delta S_0}{\delta h^{ij}}\,{,}\quad
 \pi_{\phi}=m_P^2\,\frac{\delta S_0}{\delta\phi}\,{,}
 \end{equation}
where $\pi_{ij}$ and $\pi_{\phi}$ are classical momenta
conjugates to $h^{ij}$ and $\phi$. The semi-classical region
of superspace, where $\Psi$ has an oscillating structure, is the
Lorentz one otherwise it is Euclidean. In the latter case, we have
$S=iI$ and
 \begin{equation}\label{4}
 \Psi\sim e^{-m_P^2 I}\,{,}
 \end{equation}
where $I$ is the action for the Euclidean solutions of classical field
equations ({\it istantons}). This scheme, at least at
semiclassical level, solves the problem of initial conditions.
 Given an action $S_0$, Eqs.(\ref{3}) imply $n$ free parameters
(one for each dimension of the configuration space ${\cal Q}\equiv\{h^{ij},
\phi\}$) and then $n$ first integrals of motion. The
general solution of the field equations implies $2n-1$ parameters
(one for any Hamilton equation plus the energy
constraint). As a consequence, the wave function is peaked on a subset
of the general solution and the boundary conditions on
the wave function ({\it e.g.} Harte-Hawking or  Vilenkin one) imply
initial conditions for the classical solutions.
The issue is now to search for a method capable of selecting
these constants of motion. Alternatively, can the Hartle criterion
(and the emergence of classical trajectories) be implemented by
some general approach without arbitrarily choosing regions of the
phase-space where momenta (\ref{3}) are constant?
In this review we discuss  this question considering the Hamiltonian formalism for Extended Gravity Cosmologies. We want
to show that the existence of
 Noether symmetries implies a subset of the general
solution of the WDW equation where the oscillating behaviors are selected.
Viceversa, the Hartle criterion can be  always related to
 a Noether symmetry and then to the 
classical trajectories. For {\it classical trajectories}, we mean
solutions of the standard cosmological  equations. In particular we restrict the discussion to minisuperspace models but it is clear that it could work for the complete field theory  as soon as the method could be extended to the whole superspace.
We remember that  Extended Theories of Gravity (ETGs) have recently become a sort of paradigm in the study of gravitational interaction based on corrections and enlargements of the Einstein scheme  \cite{PRnostro,physrep,physrep1}. The scheme consists, essentially, in adding higher-order curvature invariants and/or  non-minimally coupled scalar fields into dynamics resulting from the effective action of Quantum Gravity \cite{buchbinder}.
All these models are not the {\it complete theory of Quantum Gravity} but are needed as approaches toward it. 
Furthermore,  unification schemes as Superstrings, Supergravity or Grand Unified Theories, consider effective actions where non-minimal couplings to the geometry or higher-order terms  come out. 
These contributions come from   one-loop or higher-loop corrections in the high-curvature regimes \cite{birell,Vilkovisky}.
In addition,  these approaches have gained  interest in cosmology due to the fact that they {\it naturally} exhibit inflationary behaviors 
\cite{starobinsky,la,kerner,teyssandier,maeda,wands,gottloeber,sixth,aclo}.

Besides,  ETGs are going to play an interesting role to
describe also the today observed Universe. However, the energy and curvature regimes are very different with respect to the primordial epochs. 
In fact, the good
quality data of last years have made it possible to build up a more realistic  picture of the observed universe. Type Ia Supernovae
(SNeIa) \cite{SNeIa}, anisotropies in the Cosmic Microwave
Background Radiation (CMBR) \cite{CMBR}, and matter power spectrum
inferred from large galaxy surveys \cite{LSS} represent  evidences for a substantial revision of the Cosmological
Standard Model.
 In particular, the
\textit{concordance $\Lambda$CDM model} predicts that baryons
contribute only for $\sim4\%$ of the total matter\,-\,energy
budget, while the  \textit{cold dark matter} (CDM)
represents the bulk of the matter content ($\sim25\%$) and the
cosmological constant $\Lambda$ plays the role of the so called
"dark energy" ($\sim70\%$) \cite{triangle}. Although being the
best fit to a wide range of data \cite{LambdaTest}, the
$\Lambda$CDM model is severely affected by strong theoretical
shortcomings \cite{LambdaRev} that have motivated the search for
alternative models \cite{PR03}. Dark energy models mainly rely on
the implicit assumption that standard GR is the
correct theory of gravity. Nevertheless, its validity on
 large astrophysical and cosmological scales has never been
tested \cite{will}, and it is therefore conceivable that both
cosmic speed up and missing matter represent signals of a breakdown
of gravitation law as we conceive it at small scales.  This means that  one could
consider the possibility that the Hilbert\,-\,Einstein Lagrangian,
linear in the Ricci scalar $R$, could be generalized in order to cure the "dark side" 
shortcomings\footnote{Up to now, there is no final evidence that dark energy and dark matter exist at fundamental quantum level \cite{DMrew}.}.
The simplest choice is  a  function $f(R)$, that  can be encompassed
in the ETGs being a "minimal" extension of GR. It has been widely shown that $f(R)$-gravity  can easily match the dark energy and dark matter issues at cosmological and astrophysical scales 
\cite{capozzcurv,curvrev,cdtt,flanagan,allemandi,odintsovfr,mimicking,noipla,mond,jcap,Schmidt,ppnantro,arturo,matteo,salzano,nicola}.

In this review paper, we want to discuss the Hamiltonian dynamics for minisuperspace models coming from Extended Theories of Gravity.  
This problem is of fundamental interest for several reasons. First of all, it is the first step towards the Quantum Cosmology of such theories. Secondly, it allows to disentangle the further gravitational degrees of freedom defining their role into dynamics, using, in particular, conformal transformations. Finally, Hamiltonian formalism allows to select conserved quantities (Noether charges and currents) that assume a main role  in Quantum Cosmology since  can be connected to the emergence of classical universes.
It is worth noticing that the Noether Symmetry Approach \cite{nuovocim} that we are going to develop here, has been applied to different situations (see e.g.
\cite{jamil,hussain,jamil1,sanyal,modak,basilakos}) being a method extremely useful to seek for exact solutions.

The paper is organized as  follows.  In Sec.\ref{due}, we recall the Hamiltonian
formalism approach to GR and the problem of quantization starting from the well-established results of the Arnowitt-Deser-Misner (ADM) formalism. In Sec. \ref{tre} we introduce the Minisuperspace Approach to  Quantum Cosmology considering also its  limits. Sect.\ref{quattro} is devoted to
the Noether Symmetry Approach and to its connection to Quantum Cosmology.  The existence of Noether symmetries allows to select conserved momenta that acquire a straightforward relevance in view of the Hartle criterion. Extended Theories of Gravity are introduced in  in Sec.\ref{cinque} while conformal transformations are considered 
in Sec.\ref{sei}.  
In Sect.\ref{sette}, we discuss  examples of  minisuperspace models  coming from ETGs. It is easy to show that the point-like Hamiltonian coming from the Legendre transformation of the starting minisuperspace Lagrangian can be easily worked out and transformed in the corresponding WDW equation as soon as Noether symmetries are identified. Exact solutions are derived for such minisuperspace models and a singularity classification is presented according to \cite{odi2005,singularity}. Moreover, we analyze  the conformal equivalence in view of  Noether symmetries showing  how they transform under conformal transformations (Sec.\ref{sec:7}).  
 Discussion and conclusions are drawn in Sect.\ref{otto}.

\section{The Hamiltonian formulation of General Relativity and the problem of quantization}
\label{due}
Let us start with a summary of the so called canonical formulation of GR according to the so called ADM formalism \cite{halliwell,misner}.
In order to achieve  the Hamiltonian formulation of GR, 
we have to consider a $3$-surface on which the $3$-metric is $h_{ij}$,
with  matter fields defined on it.  
The $3$-manifold is embedded in a $4$-manifold whose
metric is $g_{\mu\nu}$. Following the ADM formalism, the embedding is described
by the so-called $(3+1)$ form of  $g_{\mu\nu}$, that is 
\begin{eqnarray}
ds^2 &=& g_{\mu\nu} dx^{\mu} dx^{\nu} =\nonumber\\ &&=  -\left(N^2-N_iN^i\right)
dt^2+2N_i dx^i dt+h_{ij} dx^i dx^j\,,\nonumber\\ \label{3.1}
\end{eqnarray}
where $ N $ and $ N_i $ are the lapse and shift   arbitrary functions\footnote{We adopt the 
conventions: $\mu,\nu = 0,1,2,3 $ for the Greek space-time indexes and $ i,j = 1,2,3$ for the purely spatial indexes.}, and 
describe the way in which the coordinates on a
$3$-surface is related to the preceding and following $3$-manifold.
The action is, for the moment,  the standard one of GR minimally
coupled to matter, that is 
\begin{eqnarray}
{\cal S} &=& {m_P^2 \over 16 \pi } \left[ \int _{\cal M} d^4 x \sqrt{-g}\,
(R-2\Lambda) +\right.\nonumber\\&& \left.+2 \int _{\partial {\mathcal M}} d^3 x \ \sqrt{h}\, K \right] +
{\cal S}_{(m)}\,, \label{3.2}
\end{eqnarray}
where $ K $ is the trace of the extrinsic curvature $K_{ij}$ at
the boundary $ \partial {\mathcal M} $ of the $4$-manifold ${\mathcal M}$, and is
given by
\begin{eqnarray}
K_{ij} = {1 \over 2N} \left[ - {\partial h_{ij} \over \partial t}
+ 2 D_{(i} N_{j)} \right]\,. \label{3.3}
\end{eqnarray}
 $D_i$ represents  the covariant derivative on the $3$-manifold. The action for
the matter scalar field $\phi$ is
\begin{eqnarray}
{\cal S}_{(m)} = - \frac{1}{2} \int d^4 x \sqrt{-g}\, \left[ g^{\mu\nu}
\partial_{\mu} \phi \partial_{\nu} \phi +V(\phi) \right]\,,
\label{3.4}
\end{eqnarray}
that in term of  $(3+1)$-variables, is
\begin{eqnarray}
{\cal S}&=& {m_P^2 \over 16 \pi} \int d^3x \ dt \ N \sqrt{h}\, \left[ K_{ij}
K^{ij} -K^2+\right.\nonumber\\&&\left. + {^{(3)} R} - 2 \Lambda \right] + {\cal S}_{(m)}\,. \label{3.5}
\end{eqnarray}
 The Hamiltonian\footnote{In this case, the Hamiltonian is a constraint being a sum of
constraints where the lapse  and shift functions  play the
role of Lagrange multipliers.} form
of the action is then
\begin{eqnarray}
{\cal S}=\int d^3x \ dt\left[ \dot h_{ij} \pi^{ij} + \dot \Phi \pi_{\Phi}
-N{\cal H}-N^i{\cal H}_i\right] \,,\label{3.6}
\end{eqnarray}
where $\pi^{ij}$ and $ \pi_{\Phi} $ are the momenta conjugate to
$h_{ij}$ and $ \Phi $ respectively.  
The momentum constraint is 
\begin{eqnarray}
{\cal H}_i=-2D_j \pi^j_i +{\cal H}^{(m)}_i=0\,, \label{3.7}
\end{eqnarray}
while the proper   Hamiltonian constraint is 
\begin{eqnarray}
{\cal H}&=&  {16 \pi \over m_P^2} G_{ijkl}\pi^{ij} \pi^{kl}- {m_p^2
\over 16 \pi }   \sqrt{h}\, (^{(3)}R-2 \Lambda) +{\cal H}^{(m)}=0\,,\nonumber\\&&
\label{3.8}
\end{eqnarray}
where $ G_{ijkl} $ is the so-called De Witt metric explicitly given by
\begin{eqnarray}
G_{ijkl}={1\over 2} \sqrt{h}\, \left(h_{ik} h_{jl}+h_{il}
h_{jk}-h_{ij} h_{kl}\right)\,. \label{3.9}
\end{eqnarray}
These constraints correspond to the time-space
and time-time components of the  Einstein field equations respectively. The canonical quantization
procedure is essentially based on them, as we will see below.

The so-called {\it superspace} is the framework where  classical dynamics takes place: it is the space of  all $3$-metric and matter field
configurations $(h_{ij}({\bf x}), \phi({\bf x}))$ defined on a $3$-manifold. It  is infinite
dimensional, with a finite number of coordinates $(h_{ij}({\bf x}),
\phi({\bf x}))$ at every point ${\bf x}$ of the $3$-manifold. The De Witt
metric and  metric on the matter fields determine the 
metric on superspace. It has the important property that its
signature is hyperbolic at every point ${\bf x}$ in the $3$-surface.
The signature of the De Witt metric does not  dependent on the signature
of standard space-time.

The quantum state of the
system can be  represented by a wave functional $ \Psi [ h_{ij} , \phi
] $ in the canonical quantization approach. An important characteristic of this wave
function is that is does not depend explicitly on the coordinate
time $ t $.  This is because the $3$-surfaces are compact,
and thus their intrinsic geometry
fixes almost uniquely their relative position in the
$4$-manifold.

Following the Dirac quantization, the wave function is assumed to be 
 annihilated by the classical
constraints after they have been "transformed" into operators,  that is
\begin{eqnarray}
\pi^{ij}\to -i{\delta\over \delta h_{ij}}\,,\qquad
\pi_\Phi\to-i{\delta \over \delta\phi}\,. \label{4.1}
\end{eqnarray}
The equations for $ \Psi $ are the
momentum constraint
\begin{eqnarray}
{\cal H}_i\Psi= 2i D_j {\delta \Psi \over \delta h_{ij} } + {\cal
H}^{(m)}_i \Psi = 0\,, \label{4.2}
\end{eqnarray}
and the WDW equation
\begin{eqnarray}
{\cal H}\Psi &=& \left[ - G_{ijkl} \frac{\delta}{ \delta h_{ij} }
\frac{\delta}{ \delta h_{kl} }+\right.\nonumber\\ && \left.-\sqrt{h} (^{(3)}R-2 \Lambda) +{\cal
H}^{(m)} \right] \Psi = 0\,. \label{4.3}
\end{eqnarray}

The momentum constraint implies that the wave function is the same
for configurations $(h_{ij}({\bf x}),\Phi(\bf x))$ that are related by
coordinate transformations in the $3$-surface. The
momentum constraint (\ref{4.2}) is  the quantum mechanical
expression of the invariance of the theory under $3$-dimensional
diffeomorphisms. Similarly, the WDW eq. (\ref{4.3}) represents
 the reparameterization invariance of the theory. 
Such an equation  is a second order hyperbolic
functional differential equation describing the dynamical
evolution of the wave function in superspace (the {Wave Function of the Universe}). 

Another approach to canonical quantization is to
derive the wave function by  path integrals. In this case, the wave function is an Euclidean functional integral
over a  class of $4$-metrics and matter fields, weighted
by $e^{-I}$, where $ I $ is the Euclidean action of  gravity
plus matter fields, that is
\begin{eqnarray}
\Psi [\tilde h_{ij},\tilde\phi,B]= \sum_{\cal M} \int {\cal D} g_{\mu\nu}
{\cal D} \phi e^{-I}. \label{4.6}
\end{eqnarray}
The sum is  over a given class of manifolds ${\mathcal M}$ (where  $B$ is
their boundary), and over a class of $4$-metrics $
g_{\mu\nu} $ and matter fields $\phi$ which induce the
$3$-metric $\tilde h_{ij}$ and matter field configuration $
\tilde \phi$ on the $3$-surface $B$.
If the $4$-manifold has topology $\mathbb{R} \times B$, the path
integral assumes the form
\begin{eqnarray}
\Psi[\tilde h_{ij}, \tilde \Phi,B]  &=& \int {\mathcal D} N^{\mu} \int  {\mathcal D}
h_{ij} {\mathcal D} \phi \ \delta [ \dot N^{\mu} - \chi^{\mu} ] \
\nonumber\\&&\times \Delta_{\chi} \ \exp(-I[g_{\mu\nu}, \phi] )\,. \label{4.7}
\end{eqnarray}
The delta-functional fixes the gauge condition
$\dot N^{\mu} = \chi^{\mu}$ and $\Delta_{\chi}$ is the 
Faddeev-Popov determinant.  
The $3$-metric and matter field
are integrated over a class of paths\\
$(h_{ij}({\bf x},\tau),\phi({\bf x},\tau))$  matching the argument
of  wave function on the $3$-surface $B$ where we can assume $\tau =1$, 
that is,
\begin{eqnarray}
h_{ij}({\bf x},1)= \tilde h_{ij}({\bf x}), \quad \phi({\bf x},1)=\tilde  \phi({\bf x})\,.
\label{4.8}
\end{eqnarray}
The paths are fully  specified assuming,   at the initial point, $\tau =0$.
The WDW equation and momentum
constraints, (\ref{4.2}), (\ref{4.3}) can be considered  as a quantum
 invariance under $4$-dimensional diffeomorphisms.
The wave functions generated by the
path integral (\ref{4.7}) have to satisfy the
WDW equation and momentum constraints, providing that
the path integral is invariantly constructed \cite{Halli}.
The solution of the WDW equation is generated by the
path integral and  strictly depends on how the initial and  boundary conditions 
are chosen; thus the problem  of boundary conditions is crucial in canonical quantization.

\section{The Minisuperspace Approach to Quantum Cosmology}
\label{tre}
Since the superspace is infinite dimensional and very difficult to handle, {\it minisuperspaces}  are restrictions of it where some 
symmetries are imposed a priori on the metric and the related matter fields. This approach allows to construct useful toy models that, as we said,  are not the
full Quantum Gravity but can give indications towards it.
The simplest minisuperspace consists in restricting to homogeneous and isotropic   metrics and matter
fields. 
In general, a
minisuperspace  involves  the $4$-metric
(\ref{3.1}), a lapse function  $N=N(t)$ assumed homogeneous, and the
shift function  $N^i=0$ set to zero.  One obtains
\begin{equation}
ds^2 = -N^2(t) dt^2 + h_{ij}({\bf x},t) dx^i dx^j\,. \label{5.1}
\end{equation}
 Being the $3$-metric $h_{ij} $ 
homogeneous, it is described by a finite number of
functions of $t$, $q^{\alpha}(t)$, where $\alpha=0,1,2 \cdots (n-1)$. Any Bianchi-type cosmology is suitable for such an analysis
\cite{hallow}.
The Hilbert-Einstein action  (\ref{3.2})  can be recast as 
\begin{eqnarray}
{\cal S}[h_{ij},N,N^i] &=& {m_P^2 \over 16 \pi} \int dt \ d^3x \ N
\sqrt{h} \left[ K_{ij} K^{ij} -K^2+\right. \nonumber\\&&\left.+ {^{(3)} R} - 2 \Lambda \right]\,,
\label{5.5}
\end{eqnarray}
and, in general, one gets 
\begin{eqnarray}
{\cal S}[q^{\alpha}(t), N(t)] &=& \int_0^1 dt N \left[ {1 \over 2N^2}
f_{\alpha\beta}(q) \dot q^{\alpha} \dot q^{\beta} - U(q) \right]\equiv \nonumber\\ &&\equiv \int {\cal L}
dt \label{5.6}\,,
\end{eqnarray}
where, $ f_{\alpha\beta}(q) $ is the reduced  De Witt metric,
 with signature, $ (-,+,+,+...) $. The  $t$ integration can range from $0$ to $1$ by
shifting $t$ and  scaling the lapse function. Including
matter also leads to an action
of this form, and then the  $q^{\alpha}$ functions include matter variables as
well as $3$-metric components. The $(-)$ part of the signature
corresponds to a gravitational variable.

Eq. (\ref{5.6}) has the form of  a relativistic point
particle action where the particles moves on a  $n$-dimensional  curved space-time  with a self-interaction
potential. The variation  with respect to $q^{\alpha}$, gives   the 
equations of motion
\begin{eqnarray}
{1 \over N} {d \over dt} \left( {\dot q^{\alpha} \over N} \right) + {1
\over N^2} \Gamma^{\alpha}_{\beta\gamma} \dot q^{\beta} \dot q^{\gamma} +
f^{\alpha\beta} {\partial U \over \partial q^{\beta} } = 0\,, \label{5.7}
\end{eqnarray}
where $\Gamma^{\alpha}_{\beta\gamma} $ are the  Christoffel symbols 
derived from the metric $f_{\alpha\beta}$. Varying with
respect to $N$, one gets 
\begin{eqnarray}
{1 \over 2 N^2} f_{\alpha\beta} \dot q^{\alpha} \dot q^{\beta} + U(q) = 0\,,
\label{5.8}
\end{eqnarray}
that is a constraint equation.

Eqs(\ref{5.7}) and (\ref{5.8}) describe geodesic motion in minisuperspace with a
forcing term.
The general solution of (\ref{5.7}), (\ref{5.8})
 requires $(2n-1)$ arbitrary parameters to be found.
 Eqs. (\ref{5.7}) and (\ref{5.8}) have to be equivalent,
respectively, to the $00$ and $ij$ components of the  Einstein field
equations. This statement is  not guaranteed since assuming a choice   for the
metric into the action and then taking variations to derive the
minisuperspace field equations does not necessarily yield the same
field equations.  However, it holds in several cases, in particular for Bianchi models.
In order to find the Hamiltonian,  the
canonical momenta have to be defined, that is 
\begin{eqnarray}
p_{\alpha} = {\partial {\cal L} \over \partial \dot q^{\alpha} } = f_{\alpha\beta} {\dot
q^{\beta} \over N}\,, \label{5.10}
\end{eqnarray}
and the canonical Hamiltonian is
\begin{eqnarray}
{\cal H}_c = p_{\alpha} \dot q^{\alpha} - {\cal L} = N \left[ \frac{1}{2} f^{\alpha\beta} p_{\alpha}
p_{\beta} + U(q) \right] \equiv N {\cal H}\,, \label{5.11}
\end{eqnarray}
where $f^{\alpha\beta}(q)$ is the inverse metric on minisuperspace. The
Hamiltonian form of the action is
\begin{eqnarray}
{\cal S} = \int_0^1 dt \left[ p_{\alpha} \dot q^{\alpha} - N{\cal H} \right]\,. \label{5.12}
\end{eqnarray}
This equation means that the lapse function $N$ is a Lagrange
multiplier and then the Hamiltonian constraint has to be
\begin{eqnarray}
{\cal H}(q^{\alpha},p_{\alpha}) = \frac{1}{2} f^{\alpha\beta} p_{\alpha} p_{\beta} + U(q) = 0\,.
\label{5.13}
\end{eqnarray}
This is the minisuperspace reduction equivalent to the Hamiltonian constraint of the full
theory (\ref{3.8}), integrated over the spatial hypersurfaces. Also the
momentum constraint (\ref{3.7}) is identically  satisfied.

At this point, the
canonical quantization  procedure requires  a
time-independent wave function $ \Psi(q^{\alpha}) $  that has to be
 annihilated by the quantum operator corresponding to the classical
constraint (\ref{5.13}). This fact gives rises to the the WDW equation,
\begin{eqnarray}
\hat {\cal H}(q^{\alpha}, -i {\partial \over \partial q^{\alpha} } ) \Psi(q^{\alpha})
= 0\,. \label{5.14}
\end{eqnarray}
Since the metric $f^{\alpha\beta}$ depends on $q$ there is a
factor ordering issue in (\ref{5.14}). This may be
solved by requiring that the quantization procedure is
covariant in minisuperspace,  that is unchanged by
field redefinitions of the $3$-metric and matter fields, $q^{\alpha}
\to \tilde q^{\alpha}(q^{\alpha})$. This fact restricts the possible
operator orderings to
\begin{eqnarray}
\hat {\cal H} = - \frac{1}{2} \nabla^2 + \xi {\cal R} + U(q)\,, \label{5.15}
\end{eqnarray}
where $ \nabla^2 $ and $ {\cal R} $ are the Laplacian and curvature of
the minisuperspace metric $f_{\alpha\beta}$ and $\xi$ is an arbitrary
constant.

The constant $\xi$ is  fixed as soon as  the 
minisuperspace metric is   defined by the form of the action  up to a conformal factor.
From a classical viewpoint, the constraint (\ref{5.13}) can be multiplied by an
arbitrary function of $q$, $\Omega^{-2}(q)$, and the
constraint is identical in form but has metric $\tilde f_{\alpha\beta} =
\Omega^2 f_{\alpha\beta} $ and potential $\tilde U = \Omega^{-2} U $. The
same is true in the actions (\ref{5.6}) and (\ref{5.12}) if one  rescales the lapse function, $ N \to
\tilde N = \Omega^{-2} N $. However the quantum theory should also
be insensitive to such rescaling. This is achieved if the metric
dependent part of the operator (\ref{5.15}) is  conformally
covariant, that is   the constant  $\xi$ is  the
conformal coupling
\begin{equation}
\xi = - {(n-2) \over 8 (n-1) }\,, \label{5.16}
\end{equation}
for $n \ge 2 $, where $n$ is the space dimension  \cite{halliwell,misner}.

The wave function of the universe can  be obtained also by the path integral formalism. 
However, we do not consider such an approach any more here referring the interested reader to Ref.\cite{halliwell}.

Before concluding this section, an important issue has to be addressed. It is how to interpret the probability measure  in Quantum Cosmology.
In fact, given a wave function $\Psi(q^{\alpha})$, defined in a  minisuperspace,  one
needs to  a probability measure. The question is to define a suitable  probability measure.
The WDW equation is a sort of  Klein-Gordon  equation and  a current can be defined as
\begin{eqnarray}
J = {i \over 2 } \left( \Psi^* \nabla \Psi - \Psi \nabla \Psi^*
\right)\,. \label{5.26}
\end{eqnarray}
It is conserved and satisfies the relation
\begin{eqnarray}
\nabla \cdot J = 0\,, \label{5.27}
\end{eqnarray}
thanks to the structure of the WDW equation. As in the case  of the Klein-Gordon
equation (and, in general, of hyperbolic equations), the probability derived from such a 
conserved current can be affected by   negative
probabilities. Due to this shortcoming, 
the correct measure to use should be
\begin{eqnarray}
dP = |\Psi(q^{\alpha})|^2 dV\,, \label{5.28}
\end{eqnarray}
where $dV$ is a volume element of minisuperspace. Also this assumption can be problematic since
one of the coordinates $q^{\alpha}$ is  "time", so that (\ref{5.28}) is the analogue of
interpreting $|\Psi(x,t)|^2$ in ordinary quantum mechanics as the
probability of finding the particle in the space-time
interval $dxdt$. This means that a careful discussion on the meaning of time in Quantum Cosmology has to be pursued. 
For details and alternative proposals see \cite{hartleb}. 

\section{The Noether Symmetry Approach}
\label{quattro}
As we said before, minisuperspaces are restrictions of the superspace of
geometrodynamics. They are finite-dimensional configuration spaces
on which point-like Lagrangians can be defined.
Cosmological models of physical interest can be defined on such
minisuperspaces ({\it e.g.} Bianchi models). According to the above discussion, 
a crucial role is played by the conserved currents that allow to interpret the probability 
measure and then the physical quantities obtained in Quantum Cosmology. In this context, the search for general methods to achieve conserved quantities and symmetries become  relevant. The so-called {\it Noether Symmetry Approach} \cite{nuovocim}, as we will show, can be extremely useful to this purpose.

Before taking into account specific models, let us remind some
properties of the Lie derivative and the derivation of the
Noether theorem \cite{arnold}.
 Let $L_X$ be the Lie derivative
 \begin{equation}\label{5}
 (L_X\omega)\xi=\frac{d}{dt}\omega(g_*^t\xi)\,{,}
 \end{equation}
 where $\omega$ is a differential form of $R^n$ defined on the
vector field $\xi$, $g_*^t$ is the differential of the phase flux
$\{g_t\}$ given by the vector field $X$ on a differential manifold
${\cal M}$. Let $\rho_t=\rho_{g-t}$ be the action of a one--parameter
group able to act on functions, vectors and forms on the vector
spaces $C^{\infty}({\cal M})$, $D({\cal M})$, and $\Lambda({\cal M})$
constructed
starting from ${\cal M}$.
 If $g_t$ takes the point $m\in M$ in $g_t(m)$, then $\rho_t$
takes back on $m$ the vectors and the forms defined on $g_t(m)$;
$\rho_t$ is a {\it pull back} \cite{marmo}. Then the properties
 \begin{equation}\label{6}
 \rho_{t+s}=\rho_t\rho_s\,,
 \end{equation}
 holds since
 \begin{equation}\label{7}
 g_{t+s}=g_t\circ g_s\,{.}
 \end{equation}
 On the functions $f, g\in C^{\infty}({\cal M})$ we have
 \begin{equation}\label{8}
 \rho_t(fg)=(\rho_tf)(\rho_tg)\,{;}
 \end{equation}
 on the vectors $X, Y\in D({\cal M})$,
 \begin{equation}\label{9}
 \rho_t[X, Y]=[\rho_tX, \rho_tY]\,{;}
 \end{equation}
 on the forms $\omega, \mu\in \Lambda({\cal M})$
 \begin{equation}\label{10}
 \rho_t(\omega\wedge \mu)=(\rho_t\omega)\wedge (\rho_t\mu)\,{.}
 \end{equation}
 $L_X$ is the infinitesimal generator of the one--parameter
group $\rho_t$, and, being a derivative on the algebras
$C^{\infty}({\cal M})$, $D({\cal M})$, and $\Lambda({\cal M})$,
the following properties have to hold
 \begin{eqnarray}a
 L_X(fg)&=&(L_Xf)g+f(L_Xg)\,{,} \label{11} \\
 L_X[Y, Z]&=&[L_XY, Z]+[Y, L_XZ]\,{,}\label{12}\\
 L_X(\omega\wedge\mu)&=&(L_X\omega)\wedge\mu+\omega\wedge(L_X\mu)\,{,}\label{13}
 \end{eqnarray}
 which are nothing else but the Leibniz rules for  functions,
 vectors and  differential forms, respectively. Moreover,
 \begin{eqnarray}
 L_Xf&=&Xf\,{,} \label{14} \\
 L_XY&=&adX(Y)=[X, Y]\,{,}\label{15} \\
 L_Xd\omega&=&dL_X\omega\,{,}\label{16}
 \end{eqnarray}
 where $ad$ is the self--adjoint operator and $d$ is the external
derivative by which a $p$--form becomes a $(p+1)$--form.

The discussion can be specified by considering a Lagrangian ${\cal
L}$ which is a function defined on the tangent space of
configurations $T{\cal Q}\equiv\{q_i, \dot{q}_i\}$. In this case, the vector
field $X$ is
 \begin{equation}\label{17}
 X=\alpha^i(q)\frac{\partial}{\partial q^i}+
 \dot{\alpha}^i(q)\frac{\partial}{\partial\dot{q}^i}\,{,}
 \end{equation}
 where dot means derivative with respect to $t$, and
 \begin{equation}\label{18}
 L_X{\cal L}=X{\cal L}=\alpha^i(q)\frac{\partial {\cal L}}{\partial q^i}+
 \dot{\alpha}^i(q)\frac{\partial {\cal L}}{\partial\dot{q}^i}\,{.}
 \end{equation}
 The condition
 \begin{equation}\label{19}
 L_X{\cal L}=0\,,
 \end{equation}
 implies that the phase flux is conserved along $X$: this means that a
constant of motion exists for ${\cal L}$ and the Noether theorem
holds. In fact, taking into account the Euler-Lagrange equations
 \begin{equation}\label{20}
 \frac{d}{dt}\frac{\partial {\cal L}}{\partial\dot{q}^i}-
 \frac{\partial {\cal L}}{\partial q^i}=0\,{,}
 \end{equation}
 it is easy to show that
 \begin{equation}\label{21}
 \frac{d}{dt}\left(\alpha^i\frac{\partial {\cal
 L}}{\partial\dot{q}^i}\right)=L_X{\cal L}\,{.}
 \end{equation}
 If (\ref{19}) holds,
 \begin{equation}\label{22}
 \Sigma_0=\alpha^i\frac{\partial {\cal L}}{\partial\dot{q}^i}
 \end{equation}
 is a constant of motion. Alternatively, using the Cartan
one--form
 \begin{equation}\label{23}
 \theta_{{\cal L}}\equiv\frac{\partial {\cal
 L}}{\partial\dot{q}^i}\,dq^i
 \end{equation}
 and defining the inner derivative
 \begin{equation}\label{24}
 i_X\theta_{{\cal L}}=<\theta_{{\cal L}}, X>\,{,}
 \end{equation}
 we get, as above,
 \begin{equation}\label{25}
 i_X\theta_{{\cal L}}=\Sigma_0\,,
 \end{equation}
 if condition (\ref{19}) holds.
 This representation is useful to identify cyclic variables. Using
a point transformation on vector field (\ref{17}), it is possible
to get\footnote{We indicate the quantities as
Lagrangians and  vector fields with a tilde if the
non--degenerate transformation
 $$
 Q^i=Q^i(q)\,{,}\quad \dot{Q}^i(q)=\frac{\partial Q^i}{\partial
 q^j}\,\dot{q}^j\,,
 $$
 is performed. 
 However the Jacobian determinant\\
 ${\cal J}=\parallel\partial Q^i
 /\partial q^j\parallel$ has to be non--zero.}
 \begin{equation}\label{26}
 \tilde{X}=(i_XdQ^k)\,\frac{\partial}{\partial Q^k}+
 \left[\frac{d}{dt}(i_XdQ^k)\right]\frac{\partial}{\partial\dot{Q}^k}\,{.}
 \end{equation}
 If $X$ is a symmetry also $\tilde{X}$ has this property, then it
is always possible to choose a coordinate transformation so that
 \begin{equation}\label{27}
 i_XdQ^1=1\,{,} \quad i_XdQ^i=0\,{,}\quad i\neq 1\,{,}
 \end{equation}
 and then
 \begin{equation}\label{28}
 \tilde{X}=\frac{\partial}{\partial Q^1}\,{,}\quad
 \frac{\partial\tilde{{\cal L}}}{\partial Q^1}=0\,{.}
 \end{equation}
 It is evident that $Q^1$ is the cyclic coordinate and the
dynamics can be reduced \cite{arnold}. However, the change of
coordinates is not unique and a clever choice is always important.
 Furthermore, it is possible that more symmetries are found.
In this case more cyclic variables exists. For example, if $X_1,
X_2$ are the Noether vector fields and they commute, $[X_1,
X_2]=0$, we obtain two cyclic coordinates by solving the system
 \begin{equation}\label{29}
 i_{X_1}dQ^1=1\,{,}\quad i_{X_2}dQ^2=1\,{,}
 \end{equation}
 $$
 i_{X_1}dQ^i=0\,{,}\quad i\neq 1\,{;} \quad  i_{X_2}dQ^i=0\,{,}\quad
 i\neq 2\,{.}
 $$
 If they do not commute, this procedure does not work since
commutation relations are preserved by diffeomorphisms. In this
case
 \begin{equation}\label{30}
 X_3=[X_1, X_2]\,,
 \end{equation}
 is again a symmetry since
 \begin{equation}\label{31}
 L_{X_3}{\cal L}=L_{X_1}L_{X_2}{\cal L}-L_{X_2}L_{X_1}{\cal
 L}=0\,{.}
 \end{equation}
 If $X_3$ is independent of $X_1, X_2$ we can go on until the
vector fields close the Lie algebra \cite{bianchi}.
A reduction procedure by cyclic coordinates can be implemented in
three steps: {\bf i)} we choose a symmetry and obtain new coordinates as
above. After this first reduction, we get a new Lagrangian
$\tilde{{\cal L}}$ with a cyclic coordinate; {\bf ii)} we search for new
symmetries in this new space and apply the reduction technique
until it is possible; {\bf iii)} the process stops if we select a pure
kinetic Lagrangian where all coordinates are cyclic. This case is
not very common and often it is not physically relevant. Going
back to the point of view interesting in Quantum Cosmology, any
symmetry selects a constant conjugate momentum since, by the
Euler-Lagrange equations
 \begin{equation}\label{32}
 \frac{\partial\tilde{{\cal L}}}{\partial Q^i}=0\Longleftrightarrow
 \frac{\partial\tilde{{\cal L}}}{\partial \dot{Q}^i}=\Sigma_i\,{.}
 \end{equation}
 Viceversa, the existence of a constant conjugate momentum means
that a cyclic variable has to exist. In other words, a Noether
symmetry exists.

Further remarks on the form of the Lagrangian ${\cal L}$ are
necessary at this point. We shall take into account
time--independent, non--degenerate Lagrangians ${\cal L}={\cal
L}(q^i, \dot{q}^j)$, {\it i.e.}
 \begin{equation}\label{33}
 \frac{\partial {\cal L}}{\partial t}=0\,{,}\quad
 \mbox{det} H_{ij}\equiv \mbox{det}\vert\vert\frac{\partial^2{\cal L}}
 {\partial\dot{q}^i\partial\dot{q}^j}\vert\vert \neq 0\,{,}
 \end{equation}
 where $H_{ij}$ is the Hessian.
 As in usual analytic mechanics, ${\cal L}$ can be set in the form
 \begin{equation}\label{34}
 {\cal L}=T(q^i, \dot{q}^i)-V(q^i)\,{,}
 \end{equation}
 where $T$ is a positive--defined quadratic form in the
$\dot{q}^j$ and $V(q^i)$ is a potential term. The energy function
associated with ${\cal L}$ is
 \begin{equation}\label{35}
 E_{{\cal L}}\equiv \frac{\partial {\cal L} }{\partial
 \dot{q}^i}\,\dot{q}^i-{\cal L}(q^j, \dot{q}^j)\,,
 \end{equation}
 and by the Legendre transformations
 \begin{equation}\label{36}
 {\cal H}=\pi_j\dot{q}^j-{\cal L}(q^j, \dot{q}^j)\,{,}\quad
 \pi_j=\frac{\partial {\cal L}}{\partial \dot{q}^j}\,{,}
 \end{equation}
 we get the Hamiltonian function and the conjugate momenta.
Considering again the symmetry, the condition (\ref{19}) and the
 vector field $X$ in Eq.(\ref{17}) give a homogeneous polynomial
of second degree in the velocities plus an inhomogeneous term in
the $q^j$. Due to (\ref{19}), such a polynomial has to be
identically zero and then each coefficient must be independently
zero. If $n$ is the dimension of the configuration space ({\it i.e.} the
dimension of the minisuperspace), we get $\{ 1+n(n+1)/2\}$ partial
differential equations whose solutions assign the symmetry, as we
shall see below. Such a symmetry is over--determined and, if a
solution exists, it is expressed in terms of integration constants
instead of boundary conditions.
In the Hamiltonian formalism, we have
 \begin{equation}\label{43}
 [\Sigma_j, {\cal H}]=0\,{,}\quad 1\leq j \leq m\,{,}
 \end{equation}
 as it must be for conserved momenta in quantum mechanics and the
Hamiltonian has to satisfy the relations
 \begin{equation}\label{44}
 L_{\Gamma}{\cal H}=0\,{,}
 \end{equation}
in order to obtain a Noether symmetry. The vector $\Gamma$ is defined by
\cite{marmo}
 \begin{equation}\label{45}
 \Gamma =\dot{q}^i\frac{\partial}{\partial q^i}+
 \ddot{q}^i\frac{\partial}{\partial\dot{q}^i}\,{.}
 \end{equation}
These considerations can be applied to the minisuperspace models of  Quantum Cosmology and to the
 interpretation of the wave function of the universe.
As discussed above, by a straightforward canonical quantization procedure, we have
 \begin{eqnarray}
 \pi_j & \longrightarrow & \hat{\pi}_j=-i\partial_j\,{,} \label{37} \\
 {\cal H} & \longrightarrow & \hat{\cal H}(q^j,
 -i\partial_{q^j})\,{.}\label{38a}
 \end{eqnarray}
It is well known that the Hamiltonian constraint gives the WDW
equation, so that if $\vert \Psi>$ is a {\it state} of the system
({\it i.e.} the wave function of the universe), dynamics is given by
 \begin{equation}\label{39a}
 {\cal H}\vert\Psi>=0\,{,}
 \end{equation}
 where we write the WDW equation in an operatorial way.
 If a Noether symmetry exists, the reduction procedure outlined
above can be applied and then, from (\ref{32}) and (\ref{36}), we
get
 \begin{eqnarray}
 \pi_1\equiv\frac{\partial {\cal
 L}}{\partial\dot{Q}^1}=i_{X_1}\theta_{{\cal L}}& = &
 \Sigma_1\,{,}\nonumber \\
 \pi_2\equiv\frac{\partial {\cal
 L}}{\partial\dot{Q}^2}=i_{X_2}\theta_{{\cal L}}& = &
 \Sigma_2\,{,}\label{40a} \\
\ldots\quad \ldots & & \ldots \,{,} \nonumber
 \end{eqnarray}
 depending on the number of Noether symmetries. After
quantization, we get
 \begin{eqnarray}
 -i\partial_1\vert\Psi>&=&\Sigma_1\vert\Psi>\,{,} \nonumber \\
 -i\partial_2\vert\Psi>&=&\Sigma_2\vert\Psi>\,{,} \label{41a} \\
 \ldots & & \ldots \,{,} \nonumber
 \end{eqnarray}
which are nothing else but translations along the $Q^j$ axis
singled out by corresponding symmetry. Eqs. (\ref{41a}) can be
immediately integrated and, being $\Sigma_j$ real constants, we
obtain oscillatory behaviors for $\vert \Psi>$ in the directions
of symmetries, {\it i.e.}
 \begin{equation}\label{42a}
 \vert\Psi>=\sum_{j=1}^m\, e^{i\Sigma_jQ^j}\vert \chi(Q^l)>\,{,}
 \quad m < l\leq n\,{,}
 \end{equation}
 where $m$ is the number of symmetries, $l$ are the directions
where symmetries do not exist, $n$ is the total dimension of
minisuperspace.
Viceversa, dynamics given by (\ref{39a}) can be reduced by
(\ref{41a}) if and only if it is possible to define constant
conjugate momenta as in (\ref{40a}), that is oscillatory behaviors
of a subset of solutions $\vert\Psi>$ exist only if Noether
symmetry exists for dynamics.

The $m$ symmetries give first integrals of motion and then the
possibility to select classical trajectories. In one and
two--dimensional minisuperspaces, the existence of a Noether
symmetry allows the complete solution of the problem and to get
the full semi-classical limit of Quantum Cosmology \cite{minisup}.
In conclusion, we can state that
in the semi-classical limit of quantum
cosmology, the reduction procedure of dynamics, connected to the
existence of Noether symmetries, allows to select a subset of the
solution of WDW equation where oscillatory behaviors are found.
This fact, in the framework of the Hartle interpretative
criterion of the wave
function of the universe, gives conserved momenta and trajectories
which can be interpreted as classical cosmological solutions.
Vice-versa, if a subset of the solution of WDW equation has an
oscillatory behavior, due to Eq.(\ref{19}), conserved momenta 
exist and Noether symmetries are present. In other words, {\it Noether
symmetries select classical universes}.

In what follows, we will show that such a statement holds for general classes of minisuperspaces and allows to select exact classical solutions.
In this sense, the presence of Noether symmetries is a selection criterion for classical universes.
Before this,  let us  discuss the general problem of Extended Theories of Gravity and their conformal properties. 
As we will see, most of theories of gravity can be conformally related to the Einstein one plus a suitable number of scalar fields. In this sense, the above standard minisuperspace approach works for any theory of gravity.

\section{Extending General Relativity}
\label{cinque}
In Sect.1, we discussed several issues, coming from fundamental physics, astrophysics and cosmology, that lead to take into account effective theories where the gravitational action has to be generalized with respect to the standard Hilbert-Einstein one. In Quantum Cosmology, the question of the effective action of gravity is crucial since, in general, we do not know the initial conditions from which our classical, observed universe emerged. This means that general criteria to study minisuperspace models coming from Extended Gravity are extremely relevant towards a full theory of Quantum Gravity.

In this section, without pretending to be complete, we outline the main features of higher-order and scalar-tensor gravity as examples of Extended Theories of Gravity.  For a detailed discussion, see 
\cite{physrep,physrep1}.

We will consider  two main features: 
\begin{itemize}
\item first, the geometry can 
couple non-minimally to some scalar 
field; 
\item second,  
derivatives of the metric components of order higher than second  
may appear.
\end{itemize}
 In the first case, we say that we have  
scalar-tensor  gravity, and in the second case we have  
higher order theories. Combinations of non-minimally 
coupled   and higher order terms can 
also emerge in effective 
Lagrangians, producing mixed   higher order/scalar-tensor 
gravity \cite{physrep,physrep1,review}.
 A general
class of higher-order-scalar-tensor theories in four dimensions is
 given by the action
 \begin{eqnarray} \label{V3.1}
  {\cal S}&=&\int
d^{4}x\sqrt{-g}\left[F(R,\Box R,\Box^{2}R,..\Box^kR,\phi)
 +\right.\nonumber\\&& \left.-\frac{\epsilon}{2}
g^{\mu\nu} \phi_{; \mu} \phi_{; \nu}+ {\cal L}^{(m)}\right], \end{eqnarray} where $F$ is
an unspecified function of curvature invariants and of a scalar
field $\phi$. The term ${\cal L}^{(m)}$, as above, is the minimally
coupled ordinary matter contribution, considered here as  a {\it perfect fluid}; $\epsilon$ is a constant which specifies the theory. Actually its
 values can be $\epsilon =\pm 1,0$ fixing the nature and the
 dynamics of the scalar field which can be a standard scalar
 field, a phantom field or a field without dynamics (see
 \cite{valerio,odi2005,singularity} for details).
In the metric approach, the field equations are obtained by
varying (\ref{V3.1}) with respect to  $g_{\mu\nu}$.  We get
  \begin{eqnarray} 
\label{3.2cc} G^{\mu\nu}&=&\frac{1}{{\cal
G}}\left[\kappa T^{\mu\nu}+\frac{1}{2}g^{\mu\nu} (F-{\cal G}R)+\right.\nonumber\\&& \left.+
(g^{\mu\lambda}g^{\nu\sigma}-g^{\mu\nu} g^{\lambda\sigma})
{\cal G}_{;\lambda\sigma}+\right.\nonumber\\&&+\left.\frac{1}{2}\sum_{i=1}^{k}\sum_{j=1}^{i}(g^{\mu\nu}
g^{\lambda\sigma}+
  g^{\mu\lambda} g^{\nu\sigma})(\Box^{j-i})_{;\sigma}
\right.\nonumber\\&&\left.\times\left(\Box^{i-j}\frac{\partial F}{\partial \Box^{i}R}\right)_{;\lambda}-g^{\mu\nu} g^{\lambda\sigma}\right.\nonumber\\&&\left.\times\left((\Box^{j-1}R)_{;\sigma}
\Box^{i-j}\frac{\partial F}{\partial \Box^{i}R}\right)_{;\lambda}\right]\,,
\end{eqnarray} where $G^{\mu\nu}$ is the above Einstein tensor and 

 \begin{eqnarray} 
\label{3.4gg}
  {\cal G}\equiv\sum_{j=0}^{n}\Box^{j}\left(\frac{\partial F}{\partial \Box^{j} R}
\right)\;. \end{eqnarray}

 The differential Eqs.(\ref{3.2cc}) are of order
$(2k+4)$. The stress-energy tensor is due to the kinetic part of
the scalar field and to the ordinary matter:  \begin{eqnarray}  \label{3.51}
T_{\mu\nu}=T^{(m)}_{\mu\nu}+\frac{\epsilon}{2}[\phi_{;\mu}\phi_{;\nu}-
\frac{1}{2}\phi_{;}^{\alpha}\phi_{; \alpha}]\;.  \end{eqnarray} The (possible) contribution of
a potential $V(\phi)$ is contained in the definition of $F$. From now
on, we shall indicate by a capital $F$ a Lagrangian density
containing also the contribution of a potential $V(\phi)$ and by
$F(\phi)$, $f(R)$, or $f(R,\Box R)$ a function of such fields
 without potential.

By varying with respect to the scalar field $\phi$, we obtain the
Klein-Gordon equation
  \begin{eqnarray} \label{3.62} \epsilon\Box\phi=-\frac{\partial
F}{\partial\phi}\,.  \end{eqnarray} 

The simplest extension of GR is achieved assuming, 
  \begin{eqnarray}
\label{fr}
F=f(R)\,,\qquad \epsilon=0\,, \end{eqnarray}  in the action (\ref{V3.1}).
  The standard
Hilbert-Einstein action is, of course, recovered for $f(R)=R$.
Varying with respect to $g_{\alpha\beta}$, we get 

\begin{eqnarray}\label{VAR12.32} 
f'(R)R_{\mu\nu}-\frac{f(R)}{2} \, g_{\mu\nu} =\nabla_{\mu} 
\nabla_{\nu}f'(R)-g_{\mu\nu}\Box 
f'(R) \,, \nonumber\\
\end{eqnarray}

and, after some manipulations, 

\begin{eqnarray}\label{VAR12.34}
G_{\mu\nu}&=&\frac{1}{f'(R)} \left\{
\nabla_{\mu}\nabla_{\nu} f'(R) - g_{\mu\nu}\Box f'(R)
+\right.\nonumber\\&&\left.+ g_{\mu\nu} \frac{ \left[ f(R)-f'(R) R \right]}{2}  \right\}\,,
\end{eqnarray}  
where the gravitational contribution due to higher-order terms can
be simply reinterpreted as a stress-energy tensor contribution.
This means that additional and higher-order terms in the
gravitational action act, in principle, as a stress-energy tensor,
related to the  form of $f(R)$.  Considering also the standard
perfect-fluid matter contribution, we have
 \begin{eqnarray}\label{h4}
G_{\alpha\beta}&=&\frac{1}{f'(R)}\left\{\frac{1}{2}g_{\alpha\beta}\left[f(R)-Rf'(R)\right]
+\right.\nonumber\\&&\left. f'(R)_{;\alpha\beta} -g_{\alpha\beta}\Box f'(R)\right\}+ \frac{\kappa T^{(m)}_{\alpha
\beta}}{f'(R)}=\nonumber\\&&=T^{(curv)}_{\alpha\beta}+\frac{T^{(m)}_{\alpha
\beta}}{f'(R)}\,,
\end{eqnarray}

where $T^{(curv)}_{\alpha\beta}$ is an effective stress-energy
tensor constructed by the extra curvature terms.  In the case of
GR,   $T^{(curv)}_{\alpha\beta}$ identically vanishes while the
standard, minimal coupling is recovered for the matter
contribution. The peculiar behaviour of $f(R)=R$ is  due to the
particular form of the Lagrangian itself which, even though it is
a second order Lagrangian, can be non-covariantly rewritten as the
sum of a first order  Lagrangian plus a pure divergence term. The
Hilbert-Einstein Lagrangian can be in fact recast as follows:
\begin{eqnarray}
L_{HE}&=& {\cal L}_{HE} \sqrt{-g}=\nonumber\\&&=\Big[ p^{\alpha \beta}
(\Gamma^{\rho}_{\alpha \sigma} \Gamma^{\sigma}_{\rho
\beta}-\Gamma^{\rho}_{\rho \sigma} \Gamma^{\sigma}_{\alpha
\beta})+ \nabla_\sigma (p^{\alpha \beta} {u^{\sigma}}_{\alpha
\beta}) \Big]\,,\nonumber\\
\end{eqnarray}
\noindent where:
\begin{equation}
 p^{\alpha \beta} =\sqrt{-g}  g^{\alpha \beta} = \frac{\partial {\cal{L}}}{\partial R_{\alpha \beta}}\,,
\end{equation}
$\Gamma$ is the Levi-Civita connection of $g$ and
$u^{\sigma}_{\alpha \beta}$ is a quantity constructed out with the
variation of $\Gamma$ \cite{weinberg}. Since $u^{\sigma}_{\alpha
\beta}$ is not a tensor, the above expression is not covariant;
however a standard procedure has been studied to recast covariance
in the first order theories. This clearly shows that
the field equations should consequently be second order  and the
Hilbert-Einstein Lagrangian is thus degenerate.

From the action (\ref{V3.1}), it is possible to obtain another
interesting case by choosing \begin{eqnarray} F=F(\phi)R-V(\phi)\,,\qquad \epsilon
=-1\,.\end{eqnarray} In this case, we get
\begin{equation} \label{s1} {\cal S}= \int V(\phi) \left[F(\phi) R+ \frac{1}{2} g^{\mu\nu}
\phi_{;\mu}\phi_{;\nu}- V(\phi) \right]\,, \end{equation} $V(\phi)$ and
$F(\phi)$ are generic functions describing respectively the
potential and the coupling of a scalar field $\phi$.  The
Brans-Dicke theory of gravity is a particular case of the action
(\ref{s1}) for $V(\phi)$=0. The variation with
respect to $g_{\mu\nu}$ gives the second-order field equations
\begin{eqnarray} \label{s2} 
F(\phi) G_{\mu\nu}&=& F(\phi)\left[R_{\mu\nu}-
\frac{1}{2} R _{\mu\nu} \right]=\nonumber\\&&= -\frac{1}{2} T^{\phi}_{\mu\nu}- g_{\mu\nu} \Box_{g} F(\phi)+F(\phi)_{;\mu\nu}\,, \end{eqnarray} 
 here $\Box_{g}$ is the d'Alembert
operator with respect to the metric $g$.  The energy-momentum
tensor relative to the scalar field is
\begin{equation} \label{s4} T^{\phi}_{\mu\nu}= \phi_{;\mu} \phi_{;\nu}- \frac{1}{2} g_{\mu\nu} \phi_{;\alpha} \phi_{;}^{\alpha}+g_{\mu\nu} V(\phi)\,. \end{equation} The variation with respect
to $\phi$ provides the Klein - Gordon equation, {\it i.e.} the field
equation for the scalar field:  \begin{equation} \label{s5}
\Box_{g} \phi- R F_{\phi}(\phi)+ V_{\phi}(\phi)= 0\,,
\end{equation}
where $\displaystyle{F_{\phi}(\phi)= \frac{dF(\phi)}{d\phi}}$, $\displaystyle{V_{\phi}(\phi)= \frac{dV(\phi)}{d\phi}}$. This last equation
is equivalent to the Bianchi contracted identity \cite{cqg}.

\section{Conformal Transformations}
\label{sei}

Conformal transformations are  mathematical tools that are very useful in  
Extended Theories of Gravity  in order to disentangle the further gravitational degrees of freedom coming from extended actions (see \cite{MagnanoSokolowski94,FGN98,FaraoniNadeau} for 
reviews).  In Quantum Cosmology, they  are transformations configuration spaces in minisuperspace models. The idea is to perform a conformal rescaling of the   
spacetime metric $g_{\mu\nu} \rightarrow \tilde{g}_{\mu\nu}$. 
Often a  scalar field is present in the theory and the metric 
rescaling  is accompanied by a (nonlinear) redefinition of this 
field $\phi  \rightarrow \tilde{\phi}$.  New dynamical variables
$ \left\{ \tilde{g}_{\mu\nu} , \tilde{\phi} \right\}$ are thus 
obtained. The scalar field redefinition serves the purpose of 
casting the kinetic energy density of this field in canonical 
form. The new set of variables $\left\{\tilde{g}_{\mu\nu}, 
\tilde{\phi} \right\}$ is called the {\em Einstein conformal 
frame}, while $\left\{ g_{\mu\nu}, \phi 
\right\}$ constitute the {\em 
Jordan frame}. When a scalar degree of 
freedom $\phi$ is present 
in  the theory, as in scalar tensor 
or $f(R)$ gravity, it generates the 
transformation to  the Einstein frame  in 
the sense that the 
rescaling is completely determined by a function of $\phi$. In 
principle, infinitely 
many conformal frames could be introduced, giving rise to as many 
representations of the theory. 

Let the pair  $\{{\cal M}, g_{\mu\nu}\}$ be  a spacetime, with ${\cal M}$ a smooth 
manifold of  dimension ~$n \geq 2$ and $g_{\mu\nu} $ a Lorentzian 
or 
Riemannian metric on ${\cal M}$. The point-dependent rescaling of the 
metric tensor
\begin{eqnarray}   \label{cft33}
g_{\mu\nu} \longrightarrow \tilde{g}_{\mu\nu}=\Omega^2  
g_{\mu\nu} \, ,
\end{eqnarray}
where the {\em conformal factor}
$\Omega(x) $ is a nowhere
vanishing, regular\footnote{See \cite{Bronnikov01,Bronnikov02,BronnikovShikin02} for the possibility of  continuation 
beyond singular points of the conformal factor.} 
function, is called a {\em Weyl} or {\em conformal} 
transformation. Due to this metric rescaling, the 
lengths of spacelike and timelike intervals and the norms of 
spacelike and timelike vectors are changed, while  null vectors 
and null intervals of the metric $g_{\mu\nu}$ remain 
null in the rescaled metric $\tilde{g}_{\mu\nu}$.  The light cones 
are left unchanged by the transformation (\ref{cft33}) 
and the  spacetimes  $\{{\cal M}, g_{\mu\nu}\}$ and  $\{{\cal M}, 
\tilde{g}_{\mu\nu}\} $ 
exhibit the same causal structure; the converse is also true 
\cite{Wald84}. A vector that is timelike, spacelike,
or null with respect to the  metric $g_{\mu\nu}$ has the same 
character with respect to $\tilde{g}_{\mu\nu}$, and 
{\em vice-versa}.

In the ADM  decomposition of the metric,
using the lapse function  $N$ and the shift vector $N^i$,   
the transformation properties of these quantities follow from 
Eq.~(\ref{cft33}):
\begin{equation}
\tilde{N} = \Omega \, N \,, \;\;\;\;\;\; 
\tilde{N}^i  =  N^i \,,  \;\;\;\;\;\; 
\tilde{h}_{ij} = \Omega^2  \, h_{ij} \,.
\end{equation}
The ADM mass of an asymptotically flat 
spacetime \cite{ADM62}  does not change under  the conformal
transformation  and scalar field redefinition  
\cite{CasadioGruppuso02}.
The transformation properties of various geometrical quantities 
are useful \cite{Synge55,Wald84}. 
Some of them are:
\begin{eqnarray} 
\tilde{g}^{\mu\nu}=\Omega^{-2}\, g^{\mu\nu} \,, 
\;\;\;\;\;\;\;\;\;\;
\tilde{g}=\Omega^{2n} \, g \,,
\end{eqnarray}
for the inverse metric and the metric determinant, 
\begin{eqnarray}  \label{cft35}
\tilde{\Gamma}^{\alpha}_{\beta \gamma}=
\Gamma^{\alpha}_{\beta\gamma}+\Omega^{-1}\left(
\delta^{\alpha}_{\beta} \nabla_{\gamma}\Omega +
\delta^{\alpha}_{\gamma} \nabla_{\beta} \Omega
-g_{\beta\gamma}\nabla^{\alpha} \Omega \right) \,,
\end{eqnarray}
for the Christoffel symbols, 
\begin{eqnarray}
  \widetilde{ {R_{\alpha\beta\gamma}}^{\delta}} &=&
{R_{\alpha \beta \gamma}}^{\delta}+2 \, 
\delta^{\delta}_{[\alpha} \nabla_{\beta]}\nabla_{\gamma} ( \ln 
\Omega )
+\nonumber\\&&-2g^{\delta \sigma} g_{\gamma [ \alpha}\nabla_{\beta 
]}\nabla_{\sigma}
( \ln \Omega )+\nonumber\\&&+ 2 \nabla_{[ \alpha} ( \ln \Omega ) \, \delta^{\delta}_{\beta 
]}
\nabla_{\gamma}( \ln \Omega ) 
 +\nonumber\\&&-2\nabla_{[ \alpha}( \ln \Omega ) \, g_{\beta ]
\gamma} \, g^{\delta \sigma} \, \nabla_{\sigma}( \ln \Omega ) 
+\nonumber\\
& &-  2g_{\gamma [ \alpha} \delta^{\delta}_{\beta ]} \, 
g^{\sigma \rho }
\nabla_{\sigma} ( \ln \Omega ) \, \nabla_{\rho} ( \ln \Omega ) \,,
\label{cft36}
\end{eqnarray}
for the Riemann tensor,
\begin{eqnarray}
\tilde{R}_{\alpha\beta } & =& R_{\alpha\beta }
-(n-2) \nabla_{\alpha}\nabla_{\beta }
( \ln \Omega )+\nonumber\\&&
-g_{\alpha \beta } g^{\rho\sigma } \, \nabla_{\sigma} 
\nabla_{\rho} ( \ln \Omega )+
\nonumber\\
&&+\left( n-2 \right)  \nabla_{\alpha} ( \ln \Omega ) 
\nabla_{\beta}( \ln \Omega )
+\nonumber \\
&& -\left( n-2 \right) g_{\alpha\beta }\, g^{\rho\sigma} \, 
\nabla_{\rho}( \ln \Omega )  \nabla_{\sigma}( \ln \Omega ) \,,
\label{cft36bis}
\end{eqnarray}
for the Ricci tensor, and
\begin{eqnarray}
\tilde{R} & \equiv & \tilde{g}^{\alpha\beta} 
\tilde{R}_{\alpha\beta }=
\Omega^{-2} \left[ R-2 \left( n-1 \right)  \Box \left( \ln \Omega 
\right) + \right. \nonumber \\
&&\nonumber\\
& - &\left.  \left( n-1 \right) \left( n-2 \right) \, 
\frac{g^{\alpha\beta} \nabla_{\alpha} \Omega  \, \nabla_{\beta}
\Omega}{\Omega^2}
\right] \,,  \label{cft37}
\end{eqnarray}
for the Ricci scalar.  In the case of $n=4$ space-time dimensions, 
the transformation
property of the Ricci scalar can be written as 
\begin{eqnarray} 
\tilde{R} &=& \Omega^{-2} \left[ R-\frac{6 \Box \Omega}{\Omega} 
\right]  =\nonumber \\
&&\nonumber\\
&=& \Omega^{-2}  \left[ R-\frac{12 \Box ( 
\sqrt{\Omega})}{\sqrt{\Omega}} 
+ \frac{3g^{\alpha\beta} \nabla_{\alpha} \Omega \nabla_{\beta} 
\Omega}
{\Omega^2}     \right]  \, . \label{Rconfo}
\end{eqnarray}
The Weyl tensor
${C_{\alpha\beta\gamma}}^{\delta} $ with the last index 
contravariant is conformally 
invariant,
\begin{eqnarray}  \label{cft39}
{\widetilde{ {C_{\alpha\beta\gamma}}}^{\delta}}=
{C_{\alpha\beta\gamma}}^{\delta} \, ,
\end{eqnarray}
but the same tensor with indices raised or lowered with respect to
${C_{\alpha\beta\gamma}}^{\delta}$ is not. This property 
explains the name {\em 
conformal tensor} used for 
${C_{\alpha\beta\gamma}}^{\delta}$ 
\cite{Lorentz37}.
If the original metric $g_{\alpha\beta} $ is Ricci-flat ({\em 
i.e.}, $R_{\alpha\beta}=0$), the
conformally transformed metric $\tilde{g}_{\alpha\beta}$  is not  
Eq.~(\ref{cft36bis}). In the conformally transformed 
world the  conformal 
factor $\Omega$ plays the role of an 
effective form of 
matter and 
this fact has consequences for the physical interpretation of 
the theory. A vacuum metric in the Jordan frame is not such in 
the Einstein frame,  and the interpretation 
of what is matter and 
what is gravity becomes frame-dependent 
\cite{SotiriouLiberatiFaraoni}. However, if the 
Weyl tensor
vanishes in one frame, it also vanishes 
in the conformally related frame. Conformally flat metrics are 
mapped into conformally flat metrics, a property used in 
cosmology when mapping FRW
universes (which are conformally flat) into each other. In 
particular, de Sitter spaces with  scale factor
$a(t)=a_0 \, \exp \left( H_0 t \right)$  and a constant scalar 
field as the material source are mapped into similar de Sitter
spaces.
Since, in general, tensorial quantities are not 
invariant  under 
conformal transformations, neither are the tensorial 
equations describing geometry and physics. An equation involving 
a tensor field $\psi$ is said to be {\em conformally 
invariant}
if there exists a number $w$ (the {\em conformal weight} of 
$\psi$) such that, if $\psi$ is a solution of a tensor equation 
with the metric $g_{\mu\nu}$ and the associated geometrical 
quantities, $\tilde{\psi} \equiv  \Omega^w \psi$ is a solution of 
the corresponding equation with the metric $\tilde{g}_{\mu\nu}$ 
and the associated geometry.
In addition to geometric quantities, one needs to consider 
the behavior of common 
forms of matter under conformal transformations.  It goes without 
saying that most forms of matter or fields are not conformally 
invariant: invariance under 
conformal transformations is a very 
special property. 
In general, the covariant conservation equation for a 
(symmetric) stress-energy tensor $T_{\alpha\beta}^{(m)} $ 
representing 
ordinary matter, 
\begin{eqnarray} \label{cft42}
\nabla^{\beta} \, T_{\alpha\beta}^{(m)} =0\,, 
\end{eqnarray}
is not conformally invariant \cite{Wald84}. The conformally 
transformed $\tilde{T}_{\alpha\beta}^{(m)} $ satisfies the
equation
\begin{eqnarray} \label{cft43}
\tilde{\nabla}^{\beta} \, \tilde{T}_{\alpha\beta}^{(m)} = - 
\tilde{T}^{(m)} 
  \,\tilde{\nabla}_{\alpha} \left(  \ln \Omega \right) \, 
\, \,.
\end{eqnarray}
Clearly, the conservation equation (\ref{cft42})  is conformally
invariant\index{conformal invariance}   only for a matter 
component that has vanishing trace 
$T^{(m)}$ of the energy-momentum tensor. This feature is 
associated with light-like behavior;  examples are the
electromagnetic  field and a radiative fluid with equation of 
state $P^{(m)}=\rho^{(m)}/3$.
Unless $T^{(m)}=0$, Eq.~(\ref{cft43}) describes an exchange of 
energy and momentum between matter and the scalar field $\Omega$,  
reflecting 
the fact that matter and the geometric factor $\Omega$ are 
directly coupled in the Einstein frame 
description.
Since the geodesic equation ruling the motion of free particles 
in GR can be derived from the conservation 
equation 
(\ref{cft42}) ({\em geodesic  hypothesis}), it  
follows that timelike geodesics of the original metric  
$g_{\alpha\beta}$ 
are not geodesics of the rescaled metric $\tilde{g}_{\alpha\beta} 
$ and  {\em vice-versa}. Particles  in free fall in the world  
$\left( M, 
g_{\alpha\beta} \right)$ are subject to a force proportional to the 
gradient $  \tilde{\nabla}^{\alpha} \Omega$ in the rescaled world 
$\left\{ {\cal M},  \tilde{g}_{\alpha\beta} \right\}$  (this is often 
identified as  a   fifth force acting on all massive particles 
and, therefore, 
it can be said that no massive test particles exist in the 
Einstein frame). The stress-energy tensor definition in terms 
of the  matter action $ {\cal S}^{(m)}=\int d^4x\, \sqrt{-g} \, {\cal 
L}^{(m)} $, 
\begin{eqnarray} 
\label{cft43bis}
\tilde{T}_{\alpha\beta}^{(m)}=\frac{-2}{\sqrt{ -\tilde{g} } }  \, 
\frac{  \delta
\left( \sqrt{-\tilde{g}} \, \, {\cal L}^{(m)} \right) }{\delta
\tilde{g}^{\alpha\beta} } \,,
\end{eqnarray}
together with the rescaling (\ref{cft33}) of the metric, yields
\begin{eqnarray} \label{cft43ter}
&& \tilde{T}_{\alpha\beta}^{(m)}= \Omega^{-2} \,   
T_{\alpha\beta}^{(m)} \,, \quad
 \widetilde{  { {T_{\alpha}}^{\beta}} }^{(m)} =\Omega^{-4} \, 
{{T_{\alpha}}^{\beta}}^{(m)} \,,
\quad\nonumber\\&&
{\tilde{T}}^{\alpha\beta}= \Omega^{-6} \,  
{T^{\alpha\beta}}^{(m)} \,, \quad \tilde{T}^{(m)}= \Omega^{-4} \, T^{(m)} \,.
\end{eqnarray}

The last equation makes it clear that the trace vanishes
in the Einstein  frame if and only if it 
vanishes  in the Jordan frame.

In this context, it is relevant to discuss the behavior of the Klein-Gordon equation.
In fact, the source-free 
equation $\Box \phi =0 $  in the absence
of  self-interactions  is not conformally invariant.
However, its generalization
\begin{eqnarray}  
  \label{cft40}
\Box \phi-\frac{n-2}{4(n-1)} \,R \, \phi=0\,,
\end{eqnarray}
for $n \geq 2$ is conformally invariant as pointed out in the above discussion of minisuperspace quantization procedure
\cite{Penrose64,ChernikovTagirov68,Wald84}.  It is reasonable to 
allow for the possibility  that the scalar  $\phi$ acquires a 
mass 
or other potential at high energies and, accordingly, in particle 
physics  and in cosmology  it is customary to introduce a 
potential energy density $V( \phi)$ for the Klein-Gordon scalar. 
We have already discussed how a non-minimal coupling between 
$\phi$ and the Ricci curvature arises.   The 
introduction of non-minimal coupling with $\xi \neq 0$ makes the 
theory a scalar-tensor one. 

The Klein -Gordon equation is conformally 
invariant   in four space-time  
dimensions if
$\xi=1/6$ and $V=0$ or $V=\lambda \phi^4$ \cite{Penrose64,CCJ70,Wald84}. Even a constant potential $V$, equivalent to a 
cosmological constant, corresponds to  an effective mass for the 
scalar (not to be identified with a real mass
\cite{FaraoniCooperstock95}) which breaks conformal invariance
\cite{Madsen93}. 

Although unintuitive, it is not difficult to understand why a 
quartic potential preserves conformal invariance on the 
basis of dimensional considerations. Conformal invariance  
corresponds to the absence of a characteristic length (or mass) 
scale in the physics. In general, the potential $V(\phi)$ 
contains dimensional parameters  (such as the mass $m$ in 
$V=m^2\phi^2/2$) 
but, when $V=\lambda \phi^4$, the dimension of $V$ (a mass to 
the fourth power) is carried by $\phi^4$ and the self-coupling 
constant $\lambda$ is dimensionless, {\em i.e.}, there is no 
scale associated to $V$ in this case. 
We will discuss these cases in the minisuperspace framework by considering the related  Noether symmetries.

\section{Extended Minisuperspace Models}
\label{sette}
In what follows, we shall give realizations of the above approach for minisuperspace cosmological
models derived from Extended  Theories of Gravity. As we saw in previous sections, the existence of a Noether symmetry for a given minisuperspace 
 is a sort of selection rule to recover classical behaviors in cosmic evolution.
The so called Hartle criterion to select correlated regions in the configuration space
of dynamical variables is directly connected to the presence of a Noether symmetry
and we will show that such a statement works for minisuperspace models coming from  Extended Gravity.

The approach is connected  to the search for Lagrange multipliers.
In fact, imposing  Lagrange
multipliers allow to  modify the
dynamics and select the form of  effective potentials.
By integrating the multipliers, 
solutions can be achieved. In our case, such solutions are cosmological ones. 

On the other hand,  the Lagrange multipliers are constraints capable of
reducing  dynamics in scalar-tensor and higher-order theories. Technically they
are anholonomic constraints being time-dependent. They give rise
to field equations which describe  dynamics of the further
degrees of freedom coming from Extended  Theories of Gravity. This fact is
extremely relevant to deal with  new degrees of freedom under the
standard of effective scalar fields \cite{moltiplicatori}.
Below,   we give minisuperspace examples and obtain  exact cosmological solutions for non-minimally coupled and
higher-order theories. 
In particular, we show that, by imposing Lagrange multipliers, a given  minisuperspace model becomes canonical and Noether symmetries, if exist, can be found out.
Finally, it is possible to show   that Noether symmetries allow also to classify  finite-time singularities. In other words, as soon as symmetries are broken, singularities emerge at finite.
\subsection{Scalar-Tensor Gravity Cosmologies}
%
Let us take into account a nonminimally coupled theory of gravity
of the form
 \begin{equation}\label{46a}
 {\cal S}=\int d^4x\sqrt{-g}\left[F(\phi)R+\frac{1}{2}g^{\mu\nu}
 \phi_{\mu}\phi_{\nu}-V(\phi)\right]\,{,}
 \end{equation}
 where, as said,  $F(\phi)$ and $V(\phi)$ are respectively the coupling
and the potential of a scalar field \cite{pla}. We are using
physical units $8\pi G=c=\hbar =1$, so that the standard Einstein
coupling is recovered for $F(\phi)=-1/2$.

Let us restrict to a FRW minisuperspace. The Lagrangian in
(\ref{46a}) becomes point-like, that is 
 \begin{equation}\label{47a}
 {\cal L}=6a\dot{a}^2F+6a^2\dot{a}\dot{F}-
 6kaF+a^3\left[\frac{\dot{\phi}}{2}-V\right]\,{,}
 \end{equation}
 in terms of the scale factor $a$.
The configuration space of such a Lagrangian is ${\cal Q}\equiv\{a,
\phi\}$, {\it i.e.} a bidimensional minisuperspace. A Noether
symmetry exists if (\ref{19}) holds. In this case, it has to be
 \begin{equation}\label{48a}
 X=\alpha\, \frac{\partial}{\partial a}+
 \beta\,\frac{\partial}{\partial \phi}+
 \dot{\alpha}\,\frac{\partial}{\partial\dot{a}}+
 \dot{\beta}\,\frac{\partial}{\partial\dot{\phi}}\,{,}
 \end{equation}
where $\alpha, \beta$ depend on $a, \phi$. This vector field acts on the ${\cal Q}$ minisuperspace. The system of
partial differential equation given by (\ref{19}) is
 \begin{eqnarray}a
&& F(\phi)\left[\alpha+2a\frac{\partial\alpha}{\partial
 a}\right]+
 a F'(\phi)\left[\beta+a\frac{\partial\beta}{\partial
 a}\right] =  0\,{,} \label{49a}\\
&& 3\alpha+12F'(\phi)\frac{\partial\alpha}{\partial\phi}+2a
 \frac{\partial\beta}{\partial\phi}  =  0\,{,}\label{50a} \\
&& a\beta F''(\phi)+\left[2\alpha+a\frac{\partial \alpha}{\partial a}
 +\frac{\partial\beta}{\partial\phi}\right]F'(\phi)+
 \nonumber\\&&+2\frac{\partial\alpha}{\partial\phi}F(\phi)+\frac{a^2}{6}
 \frac{\partial\beta}{\partial a} =  0\,{,} \label{51a} \\
&& [3\alpha V(\phi)+a\beta V'(\phi)]a^2+6k[\alpha F(\phi)+
 a\beta F'(\phi)]  =  0\,{.}\label{52a}\nonumber\\
 \end{eqnarray}
Prime indicates the derivative with respect to $\phi$. The
number of equations is $4$ as it has to be, being $n=2$ the $\cal Q$-dimension. Several
solutions exist for this system \cite{pla,cqg,prd}. They determine
also the form of the model since the system (\ref{49a})-(\ref{52a})
gives $\alpha, \beta$, $F(\phi)$ and $V(\phi)$. For example,
if the spatial curvature is $k=0$, a solution is
 \begin{equation}\label{53a}
 \alpha=-\frac{2}{3}p(s)\beta_0a^{s+1}\phi^{m(s)-1}\,{,}\quad
 \beta=\beta_0a^s\phi^{m(s)}\,{,}
 \end{equation}
 \begin{equation}\label{54a}
 F(\phi)=D(s)\phi^2\,{,}\quad
 V(\phi)=\lambda\phi^{2p(s)}\,{,}
 \end{equation}
 where
 \begin{eqnarray}\label{55a}
 D(s)&=&\frac{(2s+3)^2}{48(s+1)(s+2)}\,{,}\nonumber\\
 \\
 p(s)&=&\frac{3(s+1)}{2s+3}\,{,}\nonumber\\
 \\
 m(s)&=&\frac{2s^2+6s+3}{2s+3}\,{,}\nonumber\\
 \end{eqnarray}
 and $s, \lambda$ are free parameters. The change of variables
(\ref{27}) gives
 \begin{equation}\label{56a}
 w=\sigma_0a^3\phi^{2p(s)}\,{,}\quad z=\frac{3}{\beta_0\chi(s)}
 a^{-s}\phi^{1-m(s)}\,{,}
 \end{equation}
 where $\sigma_0$ is an integration constant and
 \begin{equation}\label{57a}
 \chi(s)=-\frac{6s}{2s+3}\,{.}
 \end{equation}
 Lagrangian (\ref{47a}) becomes, for $k=0$,
 \begin{equation}\label{58a}
 {\cal L}=\gamma(s)w^{s/3}\dot{z}\dot{w}-\lambda w\,{,}
 \end{equation}
 where $z$ is cyclic and
 \begin{equation}\label{59a}
 \gamma(s)=\frac{2s+3}{12\sigma_0^2(s+2)(s+1)}\,{.}
 \end{equation}
 The conjugate momenta are
 \begin{equation}\label{60a}
 \pi_z=\frac{\partial {\cal L}}{\partial \dot{z}}=\gamma(s)w^{s/3}
 \dot{w}\,{,}\quad
 \pi_w=\frac{\partial {\cal L}}{\partial \dot{w}}=\gamma(s)w^{s/3}
 \dot{z}\,{,}
 \end{equation}
 and the Hamiltonian is
 \begin{equation}\label{61a}
 \tilde{{\cal
 H}}=\frac{\pi_z\pi_w}{\gamma(s)w^{s/3}}+\lambda w\,{.}
 \end{equation}
 The Noether symmetry is given by
 \begin{equation}\label{62a}
 \pi_z=\Sigma_0\,{.}
 \end{equation}
 Quantizing Eqs. (\ref{60a}), we  get
 \begin{equation}\label{63}
 \pi\longrightarrow -i\partial_z \,{,}\quad \pi_w\longrightarrow
 -i\partial_w\,{,}
 \end{equation}
 and then the WDW equation
 \begin{equation}\label{64a}
 [(i\partial_z)(i\partial_w)+\tilde{\lambda}w^{1+s/3}]\vert\Psi>=0\,{,}
 \end{equation}
 where $\tilde{\lambda}=\gamma(s)\lambda$.
The quantum version of constraint (\ref{62a}) is
 \begin{equation}\label{65a}
 -i\partial_z\vert\Psi>=\Sigma_0\vert\Psi>\,{,}
 \end{equation}
 so that dynamics results reduced. A straightforward integration
 of Eqs. (\ref{64a}) and (\ref{65a}) gives
  \begin{equation}\label{66a}
  \vert\Psi>=\vert\Omega(w)>\vert\chi(z)>\propto
  e^{i\Sigma_0z}\,e^{-i\tilde{\lambda}w^{2+s/3}}\,{,}
  \end{equation}
which is an oscillating wave function and the Hartle criterion is
recovered. In the semi--classical limit, we have two first integrals
of motion: $\Sigma_0$ ({\it i.e.} the equation for $\pi_z$) and
$E_{{\cal L}}=0$,{\it  i.e.} the Hamiltonian (\ref{61a}) which becomes
the equation for $\pi_{w}$. Classical trajectories
in
the configuration space $\tilde{\cal Q}\equiv\{w, z\}$ are immediately
recovered
 \begin{eqnarray}
 w(t)&=&[k_1t+k_2]^{3/(s+3)}\,{,}\label{67} \\
 z(t)&=&[k_1t+k_2]^{(s+6)/(s+3)}+z_0\,{,}\label{68a}
 \end{eqnarray}
then, going back to ${\cal Q}\equiv\{a, \phi\}$,
we get the {\it classical}
cosmological behaviour
 \begin{eqnarray}
 a(t)&=&a_0(t-t_0)^{l(s)}\,{,} \label{69} \\
 \phi(t)&=&\phi_0(t-t_0)^{q(s)}\,{,} \label{70a}
 \end{eqnarray}
 where
 \begin{equation}\label{71a}
 l(s)=\frac{2s^2+9s+6}{s(s+3)}\,{,}\quad q(s)=-\frac{2s+3}{s}\,{,}
 \end{equation}
 which means that Hartle criterion selects classical universes.
Depending on the value of $s$, we get Friedman, power--law, or
pole--like behaviors.

An important remark has to be done at this point.
Noether symmetries are a useful tool also to classify singularities.
For example, for  $s=-2\pm \sqrt{2}$,   Eq.(\ref{71a})  assumes the value $l(s)=-1$. In this case, we get the solution
\begin{equation}
a(t)=\frac{a_0}{(t-t_0)}{.} \label{69s} \\
\end{equation}
This means that given values of the parameter $s$    leads  to  future singularities of the scale factor $a(t)$ where the symmetry is broken.
In order to obtain a  comprehensive  classification of   singularities, we need also to know the behavior of the
 density  $\rho$ and the pressure $p$ related to the scalar field $\phi$.  In our scalar-tensor models, they are 
\begin{equation}
\label{density1}
\rho=-\frac{1}{2D(s)\phi^2}\left[\frac{1}{2}\dot{\phi}^2+V(\phi)+12D(s)H\phi\dot{\phi}\right]\,,
\end{equation}
\begin{equation}
\label{pressure1}
p=-\frac{1}{2D(s)\phi^2}\left[\frac{1}{2}\dot{\phi}^2-V(\phi)+4D(s)\left(\phi\ddot{\phi}+2H\phi\dot{\phi}-\dot{\phi}^2\right)\right]\,,
\end{equation}
where $H=\frac{\dot{a}}{a}$ is the Hubble parameter. These equations correspond to the  minimally coupled analogue forms  
\begin{equation}
p=\frac{1}{2}\dot{\phi}^2+V(\phi)\,,
\end{equation}
\begin{equation}
p=\frac{1}{2}\dot{\phi}^2-V(\phi)\,.
\end{equation}
Clearly, also $\rho$ and $p$ are functions of  $s$ and then  it is easy to achieve the 
classification of  finite-time singularities according to the value of   parameter $s$ (see also \cite{odi2005}); that is
\begin{itemize}
\item Type I (``Big Rip'') : For $t \to t_0$, $a \to \infty$,
$\rho \to \infty$ and $|p| \to \infty$. This also includes the case of
$\rho$, $p$ being finite at $t_0$.
\item Type II (``sudden'') : For $t \to t_0$, $a \to a_0$,
$\rho \to \rho_0$ and $|p| \to \infty$
\item Type III : For $t \to t_0$, $a \to a_0$,
$\rho \to \infty$ and $|p| \to \infty$
\item Type IV : For $t \to t_0$, $a \to a_0$,
$\rho \to 0$, $|p| \to 0$ and higher derivatives of $H$ diverge.
This also includes the case when $p$ ($\rho$) or both of them tend to
some finite values while higher derivatives of $H$ diverge.
\end{itemize}
In conclusion, the presence of  Noether symmetries allows a full classification of singularities which, from another point of view,  correspond to the breaking points of symmetries.

Finally, if we take into account generic Bianchi models, the configuration
space is ${\cal Q}\equiv \{ a_1, a_2, a_3, \phi\}$ and more than one
symmetry can exist as it is shown in \cite{bianchi}. The
considerations on the oscillatory regime of the wave function of
the universe and the recovering of classical behaviors are
exactly the same.

\subsection{Fourth-Order Gravity Cosmologies}
Similar arguments work for higher--order gravity minisuperspaces. In
particular, let us consider fourth--order gravity given by the
action
 \begin{equation}\label{72a}
 {\cal S}=\int d^4x \sqrt{-g}\, f(R)\,{,}
 \end{equation}
where $f(R)$ is a generic function of scalar curvature. If
$f(R)=R+2\Lambda$, the standard second--order gravity is recovered.
We are discarding matter contributions for the sake of simplicity. Reducing the action to a
point-like, FRW one, we have to write
 \begin{equation}\label{73a}
 {\cal S}=\int dt {\cal L}(a, \dot{a}; R, \dot{R})\,{,}
 \end{equation}
where dot means derivative with respect to the cosmic time. The
scale factor $a$ and the Ricci scalar $R$ are the canonical
variables. This position could seem arbitrary since $R$ depends on
$a, \dot{a}, \ddot{a}$, but it is generally used in canonical
quantization \cite{vilenkin1,Schmidt,lambda}. The
definition of $R$ in terms of $a, \dot{a}, \ddot{a}$ introduces a
constraint which eliminates second and higher order derivatives in
action (\ref{73a}), and yields to a system of second order
differential equations in $\{a, R\}$. Action (\ref{73a}) can be
written as
 \begin{equation}\label{74a}
 {\cal S}=2\pi^2\int dt \left\{ a^3f(R)-\lambda\left [ R+6\left (
 \frac{\ddot{a}}{a}+\frac{\dot{a}^2}{a^2}+\frac{k}{a^2}\right)\right]\right\}\,{,}
 \end{equation}
where the Lagrange multiplier $\lambda$ is derived by varying
with respect to $R$. It is
 \begin{equation}\label{75a}
 \lambda=a^3f'(R)\,{.}
 \end{equation}
 Here prime means derivative with respect to $R$. To recover the
 analogy with previous scalar--tensor models, let us
 introduce the auxiliary field
 \begin{equation}\label{76a}
  p\equiv f'(R)\,{,}
 \end{equation}
 so that the Lagrangian in (\ref{74a}) becomes
 \begin{equation}\label{77a}
 {\cal L}=6a\dot{a}^2p+6a^2\dot{a}\dot{p}-6kap-a^3W(p)\,{,}
 \end{equation}
 which is of the same form of (\ref{47a}) a part the kinetic term.
This is an Helmhotz--like Lagrangian \cite{magnano} and $a, p$ are
independent fields. The potential $W(p)$ is defined as
 \begin{equation}\label{78a}
 W(p)=h(p)p-r(p)\,{,}
 \end{equation}
where
 \begin{equation}\label{rh(p)}
 r(p)=\int f'(R)dR=\int pdR=f(R)\,{,} \quad h(p)=R\,{,}
 \end{equation}
such that $h=(f')^{-1}$ is the inverse function of $f'$. The
configuration space is now ${\cal Q}\equiv\{a, p\}$ and $p$ has
the same role
of the above $\phi$. Condition (\ref{19}) is now realized by
the vector field
 \begin{equation}\label{80a}
 X=\alpha (a, p)\frac{\partial}{\partial a}+\beta(a, p)
 \frac{\partial}{\partial p}+\dot{\alpha}\frac{\partial}{\partial\dot{a}}
 +\dot{\beta}\frac{\partial}{\partial\dot{p}}\,,
 \end{equation}
 and explicitly it gives the system
 \begin{eqnarray}a
&&p\left[\alpha+2a\displaystyle\frac{\partial\alpha}{\partial a}\right]p+
a\left[\beta+a\displaystyle\frac{\partial\beta}{\partial a}\right] =
 0\,{,}\label{81}  \\
&& a^2\displaystyle \frac{\partial\alpha}{\partial p}=0\,{,} \label{82} \\
&& 2\alpha+a\displaystyle \frac{\partial\alpha}{\partial
 a}+2p\displaystyle \frac{\partial\alpha}{\partial p}+
a \frac{\partial\beta}{\partial p}=0\,{,}\label{83} \\
&& 6k[\alpha p+\beta a]+a^2[3\alpha W+
a \beta \displaystyle\frac{\partial W}{\partial
 p}]=0\,{.}\label{84a}
\end{eqnarray}
 The solution of this system, {\it i.e.} the existence of a Noether
symmetry, gives $\alpha$, $\beta$ and $W(p)$. It is satisfied for
\begin{equation}\label{85a}
  \alpha=\alpha (a)\,{,} \qquad \beta (a, p)=\beta_0 a^sp\,{,}
 \end{equation}
where $s$ is a parameter and $\beta_0$ is an integration constant.
In particular,
 \begin{eqnarray}\label{86a}
 && s=0\longrightarrow \alpha (a)=-\frac{\beta_0}{3}\, a\,{,} \quad
 \beta (p)=\beta_0\, p\,{,} \nonumber\\
 \nonumber\\ && \quad W(p)=W_0\, p\,{,} \quad
 k=0\,{,}
 \end{eqnarray}
 \begin{eqnarray}\label{87a}
 && s=-2\longrightarrow \alpha (a)=-\frac{\beta_0}{a}\,{,}\quad
 \beta (a, p) = \beta_0\, \frac{p}{a^2}\,{,} \quad\nonumber\\
 \nonumber\\&&W(p)=W_1p^3\,{,} \quad \forall \,\,\, k\,{,}
 \end{eqnarray}
 where $W_0$ and $W_1$ are constants. 
Let us discuss separately the
solutions (\ref{86a}) and (\ref{87a}).

\subsubsection{The case $s=0$}

The induced change of variables
 $\displaystyle{{\cal Q}\equiv \{a, p\}
 \longrightarrow \tilde{Q}\equiv \{w, z\}}$
 can be
 \begin{equation}\label{88a}
 w(a, p)=a^3p\,{,} \quad z(p)=\ln p\,{.}
 \end{equation}
 Lagrangian (\ref{77a}) becomes
 \begin{equation}\label{89}
 \tilde{{\cal L}}(w, \dot{w},
 \dot{z})=\dot{z}\dot{w}-2w\dot{z}^2+\frac{\dot{w}^2}{w}-3W_0w\,{.}
 \end{equation}
 and, obviously, $z$ is the cyclic variable. The conjugate momenta are
 \begin{equation}\label{90a}
 \pi_z\equiv\frac{\partial \tilde{{\cal L}}}{\partial
 \dot{z}}=\dot{w}-4\dot{z}=\Sigma_0\,{,}
 \end{equation}
 \begin{equation}\label{91a}
 \pi_w\equiv\frac{\partial \tilde{{\cal L}}}{\partial
 \dot{w}}=\dot{z}+2\frac{\dot{w}}{w}\,{.}
 \end{equation}
 and the Hamiltonian is
 \begin{equation}\label{92a}
 {\cal H}(w, \pi_w, \pi_z)=
 \pi_w\pi_z-\frac{\pi_z^2}{w}+2w\pi^2_w+6W_0w\,{.}
 \end{equation}
 By canonical quantization, reduced dynamics is given by
 \begin{equation}\label{93a}
 \left[\partial^2_z-2w^2\partial^2_w
 -w\partial_w\partial_z+6W_0w^2\right]\vert\Psi>=0\,{,}
 \end{equation}
 \begin{equation}\label{94a}
 -i\partial_{z}\vert\Psi>=\Sigma_0\,\vert\Psi>\,{.}
 \end{equation}
 However, we have done simple factor ordering considerations in the
WDW equation (\ref{93a}). Immediately, the wave function has an
oscillatory factor, being
 \begin{equation}\label{95a}
 \vert\Psi>\sim e^{i\Sigma_0z}\vert\chi(w)>\,{.}
 \end{equation}
 The function $\vert\chi>$ satisfies the Bessel differential equation
 \begin{equation}\label{96a}
 \left[w^2\partial^2_w+i\frac{\Sigma_0}{2}\,w\,\partial_w+
 \left(\frac{\Sigma_0^2}{2}-3W_0w^2\right)\right]\chi (w)=0\,{,}
 \end{equation}
 whose solutions are linear combinations of Bessel functions $Z_{\nu}(w)$
 \begin{equation}\label{97a}
 \chi (w)=w^{1/2-i\Sigma_0/4}Z_{\nu}(\lambda w)\,{,}
 \end{equation}
 where
 \begin{equation}\label{98a}
 \nu =\pm\frac{1}{4}\, \sqrt{4-9\Sigma_0^2-i4\Sigma_0}\,{,}
 \quad \lambda =\pm 9\sqrt{\frac{W_0}{2}}\,{.}
 \end{equation}
 The oscillatory regime for this component depends on the reality
 of $\nu$ and $\lambda$. The wave function of the universe, from Noether
symmetry (\ref{86a}) is then
 \begin{equation}\label{99a}
 \Psi (z, w)\sim e^{i\Sigma_0[z-(1/4)\ln w]}\,
 w^{1/2}Z_{\nu}(\lambda w)\,{.}
 \end{equation}
 For large $w$, the Bessel functions have an exponential behavior, so that the wave function (\ref{99a}) can be
written as
 \begin{equation}\label{100a}
 \Psi\sim e^{i[\Sigma_0z - (\Sigma_0/4)\ln w\pm \lambda w]}\,{.}
 \end{equation}
 Due to the oscillatory behaviour of $\Psi$,  Hartle's criterion
is immediately recovered. By identifying the exponential factor of
(\ref{100a}) with $S_0$, we can recover the conserved momenta
$\pi_z, \pi_w$ and select classical trajectories. Going back to
the old variables, we get the cosmological solutions
 \begin{equation}\label{101a}
 a(t)=a_0e^{(\lambda/6)t}\,\exp{\left\{-\frac{z_1}{3}\,
 e^{-(2\lambda/3)t}\right\}}\,{,}
 \end{equation}
 \begin{equation}\label{102a}
p(t)=p_0e^{(\lambda/6)t}\,\exp{\{z_1\,
 e^{-(2\lambda/3)t}\} }\,{,}
 \end{equation}
where $a_0, p_0$ and $z_1$ are integration constants. It is clear
that $\lambda$ plays the role of a cosmological constant and
inflationary behavior is asymptotically recovered.

\subsubsection{The case $s=-2$}
%
The new variables adapted to the foliation for the solution
(\ref{87a}) are now
 \begin{equation}\label{103a}
 w(a, p)=ap\,{,}\qquad z(a)=a^2\,{.}
 \end{equation}
 and Lagrangian (\ref{77a}) assumes the form
 \begin{equation}\label{104a}
 \tilde{{\cal L}}(w, \dot{w}, \dot{z})=3\dot{z}\dot{w}-6kw-W_1w^3\,{,}
 \end{equation}
 The conjugate momenta are
 \begin{equation}\label{105a}
 \pi_z=\frac{\partial \tilde{{\cal L}}}{\partial\dot{z}}=3\dot{w}=\Sigma_1\,{,}
 \end{equation}
 \begin{equation}\label{106a}
 \pi_w=\frac{\partial \tilde{{\cal L}}}{\partial\dot{w}}=3\dot{z}\,{.}
 \end{equation}
The Hamiltonian is given by
 \begin{equation}\label{107a}
 {\cal H}(w, \pi_w, \pi_z)=\frac{1}{3}\,
 \pi_z\pi_w+6kw+W_1w^3\,{.}
 \end{equation}
 Going over the same steps as above, the wave function of the
universe is given by
 \begin{equation}\label{108a}
 \Psi (z, w)\sim e^{i[\Sigma_1z+9kw^2+(3W_1/4)w^4]}\,{,}
 \end{equation}
 and the classical cosmological solutions are
 \begin{equation}\label{109a}
 a(t)=\pm\sqrt{h(t)}\,{,}\qquad p(t)=\pm
 \frac{c_1+(\Sigma_1/3)\,t}{\sqrt{h(t)}}\,{,}
 \end{equation}
 where
 \begin{eqnarray}\label{h(t)}
 h(t)&=&\left(\frac{W_1\Sigma_1^3}{36}\right) t^4+
 \left(\frac{W_1w_1\Sigma_1}{6}\right)
 t^3+\nonumber\\&&+\left(k\Sigma_1+\frac{W_1w_1^2\Sigma_1}{2}\right)\,
 t^2+\nonumber\\&&+w_1(6k+W_1w_1^2)\, t+z_2\,{.}
 \end{eqnarray}
 $w_1$, $z_1$ and $z_2$  are integration constants. Immediately we
 see that, for large $t$
 \begin{equation}\label{111a}
 a(t)\sim t^2\,{,}\qquad p(t)\sim \frac{1}{t}\,{.}
 \end{equation}
 which is a power-law inflationary behavior.
 An extensive discussion of Noether symmetries in $f(R)$ gravity is in \cite{antoniodefelice}.

\subsection{Higher than Fourth-Order Gravity Cosmologies}
%
Minisuperspaces which are suitable for the above analysis can be
found for higher than fourth-order theories of gravity as
\begin{equation}
{\cal S}=\int d^4x\sqrt{-g}\, f(R, \Box R)\,{.}
 \end{equation}
 In this case, the configuration space is ${\cal Q}\equiv\{a, R, \Box R\}$
 considering $\Box R$ as an independent degree of freedom
 \cite{Schmidt,lambda,sixth}. The FRW point--like Lagrangian is
 formally
 \begin{equation}\label{113a}
 {\cal L}={\cal L}(a, \dot{a}, R, \dot{R}, \Box R, \dot{(\Box R)})
 \end{equation}
 and the constraints
 \begin{equation}\label{114a}
 R=-6\left[\frac{\ddot{a}}{a}+\left(\frac{\dot{a}}{a}\right)^2+\frac{k}{a^2}\right]\,,
 \end{equation}
 \begin{equation}\label{115a}
 \Box R=\ddot{R}+3\frac{\dot{a}}{a}\dot{R}\,,
 \end{equation}
 holds. Using the above Lagrange multiplier approach, we get the
Helmholtz point--like Lagrangian
 \begin{equation}\label{116a}
 {\cal L}=6a\dot{a}^2p+6a^2\dot{a}\dot{p}-6kap-a^3\dot{h}q-a^3W(p,
 q)\,{,}
 \end{equation}
 where
 \begin{equation}\label{117a}
 p\equiv \frac{\partial f}{\partial R}\,{,}\quad
 q\equiv \frac{\partial f}{\partial \Box R}\,{,}
 \end{equation}
 \begin{equation}\label{118a}
 W(p, q)=h(p)p+g(q)q-f\,{,}
 \end{equation}
 and
 \begin{equation}\label{119a}
 h(p)=R\,{,}\quad g(q)=\Box R\,{,}\quad f=f(R,\Box R){.}
 \end{equation}
 Now the minisuperspace is $3$-dimensional but, again, the
Noether symmetries can be recovered. Cases of physical interest
\cite{sixth} are
 \begin{equation}\label{120a}
 f(R, \Box R)=F_0R+F_1R^2+F_2R\Box R\,,
 \end{equation}
 \begin{equation}\label{121a}
 f(R, \Box R)=F_0R+F_1\sqrt{R\Box R}\,{,}
 \end{equation}
 discussed in details in \cite{oneloop}.  Also here the existence
of the symmetry selects the form of the model and allows to reduce the
dynamics. Once it is identified, we can perform the change of
variables induced by foliation using Eqs. (\ref{29}), if a
symmetry is present, is two symmetries are
present. In both cases,
 \begin{equation}\label{122a}
 {\cal Q}\equiv \{a, R, \Box R\}\longrightarrow \tilde{\cal Q}\equiv
 \{z, u, w\}\,{,}
 \end{equation}
 where one or two variables are cyclic in Lagrangian (\ref{116a}).
Taking into account, for example, the case (\ref{121a}), we get
 \begin{equation}\label{123a}
 \tilde{{\cal
 L}}=3[w\dot{w}^2-kw]-F_1\left[3w\dot{w}^2u+3w^2\dot{w}\dot{u}+
 \frac{w^3\dot{z}\dot{u}}{2u^2}-3kwu\right]\,{,}
 \end{equation}
 where we assume $F_0=-1/2$, the standard Einstein coupling, $z$ is the
cyclic variable and
 \begin{equation}\label{124a}
 z=R\,{,}\quad u=\sqrt{\frac{\Box R}{R}}\,{,}\quad w=a\,{.}
 \end{equation}
 The conserved quantity is
 \begin{equation}\label{125a}
 \Sigma_0=\frac{w^3\dot{u}}{2u^2}\,{.}
 \end{equation}
 Using the canonical procedure of quantization and deriving the
WDW equation from (\ref{123a}), the wave function of the universe
is
 \begin{equation}\label{126a}
\vert \Psi>\sim e^{i\Sigma_0z}\vert\chi(u)>\vert\Theta(w)>\,{,}
 \end{equation}
 where $\chi(u)$ and $\Theta(w)$ are combinations of Bessel
functions. The oscillatory subset of the solution is evident and
the Hartle criterion is recovered. In the semi--classical limit,
using the conserved momentum (\ref{125a}), we obtain the
cosmological behaviours
 \begin{equation}\label{127a}
 a(t)=a_0t\,{,} \quad a(t)=a_0t^{1/2}\,{,}\quad
 a(t)=a_0e^{k_0t}\,{,}
 \end{equation}
 depending on the choice of boundary conditions.
 
 However, the above considerations of singularities at finite time holds also for higher-order theories of gravity. Also in these cases, they can be classified according to the presence of Noether symmetries.

\section{Noether Symmetries and Conformal Transformations}
\label{sec:7}
%
To conclude the discussion, 
we want now to analyze the compatibility between the conformal
transformation connecting Jordan frame and Einstein frame, and the
presence of Noether symmetries. In other words, we want to analyze how minisuperspace models, 
derived by the Noether Symmetry Approach, behave under the action of conformal transformation.  

Since the existence of a Noether symmetry implies the existence of a
vector field $X$ along which $L_{X}{\cal L}= 0$, this happens if the Lie
derivative of the Lagrangian along a vector field is preserved. 
  Let us define a vector field $X$ (in time $\eta$)  acting on the minisuperspace $\cal Q$ in
the Einstein frame and in  the Jordan frame. We want to investigate how this vector field acts on one of the above  point-like Lagrangians under a conformal transformation. Let us consider the minisuperspace model (\ref{46a}).
 
Thus, a given
lift-vector field of the form \cite{marmo}
\begin{eqnarray}
\label{38b}
X_\eta= \alpha \frac{\partial}{\partial a}+ \beta \frac{\partial}{\partial \phi}+ \alpha^{'} 
\frac{\partial}{\partial a'}+ \beta^{'} \frac{\partial}{\partial \phi'}\,,
\end{eqnarray}
where $\alpha= \alpha(a,~\phi)$, $\beta= \beta(a,~\phi)$, corresponds, under conformal
transformations, to the lift-vector field on the configuration space
$({\bar a},~{\bar \phi})$
\begin{eqnarray}
\label{39b}
\bar{X}_\eta= \bar{\alpha} \frac{\partial}{\partial \bar a}+ \bar{\beta}
\frac{\partial}{\partial \bar \phi}+ \bar{\alpha}^{'} \frac{\partial}{\partial \bar a'}+
\bar{\beta}^{'} \frac{\partial}{\partial \bar \phi'} \,,
\end{eqnarray}
where $\bar{\alpha}= \bar{\alpha}(\bar a,~\bar \phi)$, $\bar{\beta}= 
\bar{\beta}(\bar a,~\bar \phi)$ are connected to $\alpha= \alpha(a,~\phi)$, $\beta=
\beta(a,~\phi)$ through a Jacobian matrix. Here the prime means
the derivative with respect to the time $\eta$. The Lie derivative of the Lagrangian 
$ {\cal L}_\eta$ along the vector field $X_\eta$ corresponds then to the Lie 
derivative of $\bar { {\cal L}}_\eta$ along ${\bar{X}}_\eta$ \cite{tre-grazie}
\begin{eqnarray}
\label{40b}
 {\cal L}_{X_\eta}  L_\eta=  L_{\bar{X}_\eta} \bar {\cal L}_\eta.
\end{eqnarray}
Therefore,   if $X_\eta$ is a Noether vector field relative to ${\cal L}_\eta$ 
one has
\begin{eqnarray}
\label{41b}
L_{X_\eta} {\cal L}_\eta= 0\,,
\end{eqnarray}
and, from (\ref{40b}), $\bar{X}_\eta$ is a Noether vector field 
relative to $\bar {\cal  L}_\eta$.
The choice of $\eta$
as time-coordinate is convenient from a formal point of view, but in
order to analyse the phenomenology relative to a given model and to
obtain then quantities comparable with the observational data, the
appropriate choice of time-coordinate is the cosmic time $t$. The
problem with the cosmic time is that it is not preserved by the
conformal transformation. Thus the
conformal transformation we are considering does not take simply the form of
a ``coordinate transformation'' on the phase space $\{ a,~\phi \}$,
therefore its compatibility with the presence of a Noether symmetry
cannot be easily verified. Of course it must hold also under such a
choice of time--coordinate.

We do not decide to verify such a compatibility directly. Rather, we
analyse how the Lie derivative $ L_{X_\eta} {\cal L}_\eta$ in the Jordan
frame is transformed under the time transformation which
connects $t$ with $\eta$ \cite{noether}.
The explicit expression of $L_{X_\eta} {\cal L}_\eta$ is given by
\begin{eqnarray}
\label{42b}
L_{X_\eta} {\cal L}_\eta &=& 6 \left[ 2 F \frac{\partial \alpha}{\partial a}+
\left( \beta+ a \frac{\partial \beta}{\partial a}\right) F_{\phi}\right] a^{'2}+\nonumber\\&&+ a
\left[ \alpha+ 6 F_{\phi} \frac{\partial \alpha}{\partial \phi}+ a \frac{\partial \beta}{\partial
\phi}\right] \phi^{'2}+\nonumber\\ &&
+6 \left[ a \beta F_{\phi\phi}+ \left( \alpha+ a \frac{\partial \alpha}{\partial
a}+ a \frac{\partial \beta}{\partial \phi}\right) F_{\phi}+\right.\nonumber\\&&\left.+ 2 F \frac{\partial \alpha}{\partial \phi}+
\frac{a^2}{6} \frac{\partial \beta}{\partial a}\right] a' \phi'+\nonumber\\&&
 -a^3 (4 \alpha V+ a \beta V_{\phi})- 6 a (2 F \alpha+ F_\phi a \beta) 
\kappa\,,\nonumber\\
\end{eqnarray}
in which we have taken into account that
\begin{eqnarray}
\label{43b}
\alpha^{'}= \frac{\partial \alpha}{\partial a} a'+ \frac{\partial \alpha}{\partial \phi} \phi';~~~ 
\beta^{'}= \frac{\partial \beta}{\partial a} a'+ \frac{\partial \beta}{\partial \phi} \phi'.
\end{eqnarray}
Eq. (\ref{42b}), under the transformation becomes
\begin{eqnarray}
\label{44b}
 L_{X_\eta} {\cal L}_\eta &=&6 \left[ 2 F \frac{\partial \alpha}{\partial a}+
\left( \beta+ a \frac{\partial \beta}{\partial a}\right) F_{\phi}\right] a^2 {\dot a}^2+\nonumber\\&&+ a
\left[ \alpha+ 6 F_{\phi} \frac{\partial \alpha}{\partial \phi}+ a \frac{\partial \beta}{\partial
\phi}\right] a^2 {\dot \phi}^2+\nonumber\\&& 
 +6 \left[ a \beta F_{\phi\phi}+ \left( \alpha+ a \frac{\partial \alpha}{\partial
a}+ a \frac{\partial \beta}{\partial \phi}\right)F_{\phi}+\right.\nonumber\\&&\left.+ 2 F \frac{\partial \alpha}{\partial \phi}+
\frac{a^2}{6} \frac{\partial \beta}{\partial a}\right] a^2 {\dot a} {\dot \phi}+\nonumber\\&&-a^3 (4 \alpha V+ a \beta V_{\phi})- 6 a (2 F \alpha+ F_{\phi} a \beta)
\kappa\,,\nonumber\\
\end{eqnarray}
which can be written as
\begin{eqnarray}
\label{45b}
L_{X_\eta} {\cal L}_\eta & = & 6 a \left[ \alpha F+ 2 a F \frac{\partial
\alpha}{\partial a}+ a \left( \beta+ a \frac{\partial \beta}{\partial a}\right) F_{\phi}\right]
{\dot a}^2+\nonumber\\&&+ a^3 \left[ \frac{3}{2} \alpha+ 6 F_{\phi} \frac{\partial \alpha}{\partial \phi}+ a
\frac{\partial \beta}{\partial \phi}\right] {\dot \phi}^2 +\nonumber\\&&+6 a^2 \left[ a \beta F_{\phi\phi}+ \left( 2 \alpha+ a \frac{\partial
\alpha}{\partial a}+ a \frac{\partial \beta}{\partial \phi}\right) F_\phi+\right.\nonumber\\ &&\left.+ 2 F \frac{\partial
\alpha}{\partial \phi}+ \frac{a^2}{6} \frac{\partial \beta}{\partial a}\right] {\dot a} {\dot \phi}-a^3 (3 \alpha V+ a \beta V_\phi)+\nonumber\\&&- 6 a (F \alpha+ F_\phi a \beta) \kappa-6 \alpha F a {\dot a}^2- \frac{1}{2} \alpha a^3 {\dot \phi}^2+\nonumber\\&&- 6 \alpha F_\phi a^2 {\dot a}
{\dot \phi}- 6 \alpha F a \kappa- \alpha a^3 V. 
\end{eqnarray}
The Lie derivative of ${\cal L}_t$, along the lift--vector 
field of the form
\begin{eqnarray}
\label{46b}
X_t= \alpha \frac{\partial}{\partial a}+ \beta \frac{\partial}{\partial \phi}+ \dot{\alpha}
\frac{\partial}{\partial {\dot a}}+ \dot{\beta} \frac{\partial}{\partial {\dot \phi}},
\end{eqnarray}
is given by
\begin{eqnarray}
\label{47b}
 L_{X_t} {\cal L}_t &=& 6 \left[ \alpha F+ 2 a F \frac{\partial \alpha}{\partial a}+ a
\left( \beta+ a \frac{\partial \beta}{\partial a}\right) F_\phi\right] {\dot a}^2+\nonumber\\&&+ a^2
\left[ \frac{3}{2} \alpha+ 6 F_\phi \frac{\partial \alpha}{\partial \phi}+ a \frac{\partial
\beta}{\partial \phi}\right] {\dot \phi}^2+\nonumber\\&&+6 a \left[ a \beta F_{\phi\phi}+ \left( 2 \alpha+ a \frac{\partial
\alpha}{\partial a}+ a \frac{\partial \beta}{\partial \phi}\right) F_\phi+\right.\nonumber\\&&\left.+ 2 F \frac{\partial
\alpha}{\partial \phi}+ \frac{a^2}{6} \frac{\partial \beta}{\partial a}\right] {\dot a} {\dot \phi} -a^2 (3 \alpha V+ a \beta V_\phi)+\nonumber\\&&- 6 (F \alpha+ F_\phi a \beta) \kappa,
\end{eqnarray}
in which we have taken into account that 
\begin{eqnarray}
\label{48b}
\dot{\alpha}= \frac{\partial \alpha}{\partial a} {\dot a}+ \frac{\partial \alpha}{\partial \phi} {\dot \phi};~~~ 
\dot{\beta}= \frac{\partial \beta}{\partial a} {\dot a}+ \frac{\partial \beta}{\partial \phi} {\dot \phi}.
\end{eqnarray}
We remind that the dot means the derivative with respect to $t$. 

Comparing (\ref{45b}) with (\ref{47b}), we obtain that,
 the Lie derivative $ L_{X_\eta} {\cal L}_\eta$ 
becomes
\begin{eqnarray}
\label{49b}
{L}_{X_\eta} {\cal L}_\eta= a {L}_{X_t} {\cal L}_t- ( { L}_{X_t} a) E_t,
\end{eqnarray}
being $ {\cal L}_{X_t} a= \alpha$ where $E_t$ is the energy function.

It can be seen that the same relation as (\ref{49b}) holds in the 
Einstein frame, that is 
\begin{eqnarray}
\label{51b}
 {\cal L}_{{\bar X}_{\eta}} {\bar L}_\eta=
  {\bar a}  {\cal L}_{{\bar X}_{\bar t}} {\bar L}_{\bar t}- 
( {\cal L}_{{\bar X}_{\bar t}} {\bar a}) {\bar E}_{\bar t},
\end{eqnarray}
with  obvious meaning of $\bar{X}_{\bar t}$ and ${\bar E}_{\bar t}$.
This implies that, if $X_\eta$ is a Noether vector field relative to
${\cal L}_\eta$, that is, if (\ref{41b}) holds, the corresponding vector field
$X_t$ is such that 
\begin{eqnarray}
\label{52b}
L_{X_t} {\cal L}_t- \frac{\L_{X_t} a}{a} E_t= 0.
\end{eqnarray}
It means also that, when the cosmic time is taken as
time--coordinate, the conformal transformation preserves the
expression given by the righthand side of (\ref{49b}) and not the Lie
derivative along a given vector field $X_t$. Relation (\ref{52b})
represents a more general way to express the presence of a first
integral for the Lagrangian $L_t$; associated to (\ref{52b}), we have
the conserved quantity \cite{logan} 
\begin{eqnarray}
\label{53b}
\frac{\partial {\cal L}_t}{\partial {\dot a}} \alpha+ \frac{\partial {\cal L}_t}{\partial {\dot \phi}} \beta -E_t \int
\frac{L_{X_t} a}{a} dt= const,
\end{eqnarray}
which, of course, holds on the solutions of the Euler--Lagrange 
equations. The vector field $X_t$ verifying (\ref{52b}) can thus be 
seen as a generalized Noether vector field and the conformal 
transformation (\ref{21}) preserves this generalized symmetry. That 
is, if $X_t$ is a Noether vector field, in the sense of (\ref{52b}), 
relative to ${\cal L}_t$ then $\bar{X}_{\bar t}$ is a Noether vector field relative 
to ${\bar {\cal L}}_{\bar t}$ in the same sense, that is 
\begin{eqnarray}
\label{54b}
{ L}_{\bar{X}_{\bar t}} {\cal {\bar L}}_{\bar t}- \frac{{ L}_{\bar{X}_t} {\bar a}}{\bar a} \bar{E}_t= 0.
\end{eqnarray}
In terms of  conformal time, the first integral relative to (\ref{41b}) 
for the Lagrangian ${\cal L}_\eta$ is given by
\begin{eqnarray}
\label{55b}
\frac{\partial {\cal L}_\eta}{\partial a'} \alpha+ \frac{\partial {\cal L}_\eta}{\partial \phi'} \beta=
const.
\end{eqnarray}
We see that the expression (\ref{55b}) corresponds to (\ref{53b}) under 
the transformation, except for a term in the energy 
function. In fact, Eq.(\ref{55b}) explicitly written, is
\begin{eqnarray}
\label{56b}
(12 F \alpha'+ 6 F_\phi a \phi') \alpha+ (6 F_\phi a \alpha'+ a^2 \phi') \beta= const,\nonumber\\
\end{eqnarray}
while (\ref{53b}) is
\begin{eqnarray}
\label{57b}
&&-E_t \int \frac{\alpha}{a} dt+ (12 F a {\dot a}+6 F_\phi a^2 {\dot \phi}) \alpha+\nonumber\\&&+ (6 
F_\phi a^2 {\dot a}+ a^3 {\dot \phi}) \beta= const.
\end{eqnarray}
Taking into account that,  the
expressions of  energy function, in the conformal time and in the
cosmic time, are related by
\begin{eqnarray}
\label{31b}
E_\eta= a E_t,
\end{eqnarray}
we have that (\ref{57b}), becomes
\begin{eqnarray}
\label{58b}
&&-\frac{E_\eta}{a} \int \alpha d\eta+ (12 F a'+ 6 F_\phi a \phi') \alpha+\nonumber\\&& +(6 
F_\phi a a'+ a \phi') \beta= const
\end{eqnarray}
which coincides with (\ref{56b}) except for the term in $E_\eta$, but
$E_\eta= 0$ corresponds to the first order Einstein equation,
expressed in the time $\eta$, therefore,
there is equivalence between the two formulations \cite{noether}. 
As already said, some  authors have formulated the existence of
a Noether vector field imposing 
\begin{eqnarray}
\label{59b}
{ L}_{X_t} {\cal L}_t= 0\,,
\end{eqnarray}
using the cosmic time as time-coordinate; condition (\ref{59b}),
after the analysis we have done till now, turns out to be less general
than (\ref{54b}). By the way, condition (\ref{59b}) has the interesting
property that it implies the possibility to define some new
coordinates on the configuration space $\{ a,~\phi\}$, such that the
Lagrangian has a cyclic coordinate \cite{nuovocim,marmo},
reducing in this way the Euler-Lagrange equations. In fact, one can
always define new coordinate, say $\{z,~w\}$, in the configuration space
of the Lagrangian, such that the lift-vector field assumes the form
${\displaystyle X_t= \frac{\partial }{ \partial z}}$, so that one has ${\displaystyle { L}_{X_t} {\cal L}= \frac{\partial
{\cal L}}{\partial z}}$; in case Eq.(\ref{59b}) holds, one has that $z$ is cyclic.
In the generalized case we are considering, it is no longer possible
to get this behavior, since ${L}_{X_t} {\cal L}_t \neq 0$ and consequently $z$
is no longer cyclic. In this case, one has to use the first integral
(\ref{53b}) together with the relation on the energy function to reduce
the Euler-Lagrange equations.

This problem corresponds, in the Einstein frame, to
\begin{eqnarray}
\label{60b}
-{\bar E}_{\bar t} \int \frac{{ L}_{\bar{X}_{\bar t}} {\bar a}}{\bar a} d{\bar t}+
 \frac{\partial{\bar {\cal L}}_{\bar t}} {\partial {\dot{\bar a}}} \bar{\alpha}+ 
 \frac{\partial {\cal \bar L }_{\bar t}}{\partial{ \dot{ \bar \phi}}}
\bar{\beta}= const. 
\end{eqnarray}
Thus, finding the solutions of some cosmological model by the 
existence of a Noether symmetry (and therefore fixing the class of 
model compatible with it) in the Einstein frame, one gets, via the 
conformal transformation, the solutions to the
class of models in the Jordan frame corresponding to the one given in
the Einstein frame. This result is extremely important in minisuperspace Quantum Cosmology since allow to select all the equivalent initial (boundary) conditions.

Let us discuss a significant example.
A  solvable cosmological model in the Einstein frame 
is the model where the dynamical potential of the scalar
field is constant and the spatial curvature is zero. The Lagrangian is
given by 
\begin{eqnarray}
\label{61b}
{\cal \bar L}_{\bar t}= -3 {\bar a} {\dot{\bar a}}^2+ \frac{1}{2}  {\bar a}^3 {\dot{\bar \phi}}^2- {\bar a}^3 
\Lambda.
\end{eqnarray}
We have that such a model in the
Einstein frame corresponds, in the Jordan frame, to the class of
models with (arbitrarly given) coupling $F$ and potential $V$
connected by the relation 
\begin{eqnarray}
\label{71b}
\frac{V}{4 F^2}= \Lambda.
\end{eqnarray}
We can thus fix the potential $V$ and obtain from (\ref{71b}) the
corresponding coupling. Once we have the solutions relative to the
constant potential in the Einstein frame, we can obtain the solutions
of all the non-minimally coupled models in the class defined by
relation (\ref{71b}). 

Let us consider the case
\begin{eqnarray}
\label{72b}
V= \lambda \phi^4,~~~ \lambda> 0\,,
\end{eqnarray}
which correspond to a ``chaotic inflationary'' potential. The
corresponding coupling is quadratic in $\phi$, that is
\begin{eqnarray}
\label{73b}
F= k_0 \phi^2\,,
\end{eqnarray}
in which
\begin{eqnarray}
\label{74b}
k_0= -\frac{1}{2} \sqrt{\frac{\lambda}{\Lambda}}.
\end{eqnarray}
 We get the conformal 
transformation through which we obtain the solutions in the Jordan 
frame, {\it i.e.}
\begin{eqnarray}
\label{75b}
 a & =& \frac{\bar a}{\phi \sqrt{-2 k_0}}\,, \nonumber\\
 \nonumber\\
  d \phi &=& \phi \sqrt{\frac{2 k_0}{12 k_0 -1}}\, d{\bar \phi}\,, \nonumber\\
  \nonumber\\
      dt &=& \frac{d{\bar t}}{\phi \sqrt{-2 k_0}}. \nonumber\\
\end{eqnarray}
As we can see from these relations, it has to be $k_0< 0$. Integrating
the second of (\ref{75b}), we have $\phi$ in terms of $ {\bar \phi}$
\begin{eqnarray}
\label{76b}
\phi= \alpha_0 e^{\sqrt{\frac{2 k_0}{12 k_0- 1}} \, {\bar \phi}}.
\end{eqnarray}

 We want now to consider the aspects
connected with the point of view of the Noether symmetries. It is
possible to show that, in the context of generalized Noether
symmetries, the nonstandard coupled model with quartic potential and
negative quadratic coupling admits a Noether symmetry, while such a
result has not been found in the previous analysis of Noether
symmetries (see  \cite{noether,nuovocim}). The system of equations
for the Noether vector field obtained from (\ref{54b}) is given by
\begin{eqnarray}
\label{87b}
\frac{\partial \bar{\alpha}}{\partial \bar a}= 0\,, \\
\nonumber\\
\bar\alpha+ {\bar a} \frac{\partial \bar{\beta}}{\partial \bar \phi}= 0\,, \\
\\
 6 \frac{\partial \bar{\alpha}}{\partial \bar \phi}- {\bar a}^2 \frac{\partial \bar{\beta}}{\partial 
\bar a}=0\,,\\
\\
4 \alpha \bar V+ \bar a \bar{\beta} {\bar V}_{\bar \phi}= 0.
\end{eqnarray}
Substituting $\bar V= \Lambda$ in the fourth of (\ref{87b}), one gets
$\bar{\alpha}= 0$; from the second one gets $\bar{\beta}= const$; the first
and the third turn out to be identically verified. It is immediate to
see that the Lagrangian (\ref{61b}) presents a Noether symmetry, since
it does not depend on $\bar \phi$; being, in this particular case,
${L}_{\bar{X}_{\bar t}} {\bar a}= \bar{\alpha}= 0$.  This result is compatible with the presence of a cyclic coordinate in the
Lagrangian. Performing the conformal transformation given by
(\ref{75b}) on the Noether vector field 
\begin{eqnarray}
\label{88b}
\bar{\alpha}= 0\,, \quad
\bar{\beta}= \bar{\beta}_0\,, 
\end{eqnarray}
we have
\begin{eqnarray}
\label{89b}
 \alpha= -a \sqrt{\frac{2 k_0}{12 k_0- 1}}\, \bar{\beta}_0\,, \quad
 \beta= \phi \sqrt{\frac{2 k_0}{12 k_0- 1}}\, \bar{\beta}_0.\nonumber\\
\end{eqnarray}
This means that  (\ref{89b}) is a Noether
vector field relative to the corresponding Lagrangian in the Jordan
frame, with potential given by (\ref{72b}) and coupling given by
(\ref{73b}). It is easy to verify that (\ref{52b}) holds.

In conclusion, the Noether Symmetry Approach is compatible with conformal transformations and allows to relate classes of conformally equivalent minisuperspace models

 \section{Discussion and Conclusions}
 \label{otto}
%
The purpose of this  paper has been to outline
the canonical Hamiltonian approach to Quantum Cosmology taking into account minisuperspace models coming from Extended Theories of Gravity.
After a quick summary of the Hamiltonian formulation of GR and  the problem of canonical quantization in the ADM formalism, we discussed the minisuperspace approach to Quantum Cosmology. This one does not give  a satisfactory solution to the full Quantum Gravity problem, however it is a useful scheme to set the problem of boundary conditions from which should emerge {\it classical universes}, that is cosmological dynamical models that could be reasonably {\it observed} with standard astrophysical tools. A main role in this approach is played by the identification of conserved quantities that give rise to peaked behaviors in the wave function of the universe. Such a function is the solution of the WDW equation, the corresponding of Schr\"odinger equation in Quantum Cosmology. 

Peaked behaviors means correlations among variables and then the possibility to obtain classical universes according to the Hartle interpretative criterion. These conserved quantities can naturally be related to the Noether symmetries of the theory. The existence of symmetries depends, in several cases, by the identification of suitable Lagrange multipliers  that allow to recast the point-like Lagrangian of the given minisuperspace model in a canonical form. In this sense, the Noether symmetries can be considered as "constraints" of the theory that allow to reduce the dynamics and  recover  classical solutions.  The emergence of singularities at finite for such solutions means that symmetries are broken for certain values of the parameters.

Reversing the argument, if the wave function of the universe is
related to the probability to get a  classical cosmological solution, the
existence of Noether symmetries tell us when the Hartle criterion
works.

Some remarks are necessary at this point. First of all, we have to
stress that the wave function is {\it only} related to the
probability to get a certain behavior but it is not the
probability amplitude since, till now, Quantum Cosmology is not a
unitary theory. Furthermore, the Hartle criterion works in the
context of an Everett-type interpretation of Quantum Cosmology
\cite{everett,finkelstein} which assumes the ideas that the
universe branches into a large number of copies of itself whenever
a measurement is made. This point of view is called {\it Many
Worlds} interpretation of Quantum Cosmology.
Such an interpretation is just one way
of thinking and gives a formulation of Quantum Mechanics designed
to deal with correlations internal to individual, isolated systems.
The Hartle criterion gives an operative interpretation of such
correlations. In particular, if the wave function is {\it strongly
peaked} in some region of configuration space, we predict that we
will observe the correlations which characterize that region. On
the other hand, if the wave function is {\it smooth} in some
region, we predict that correlations which characterize that
region are precluded to the observations.
If the wave function is neither peaked nor smooth, no predictions
are possible from observations. In other words, we can read the
{\it correlation} of some region of minisuperspace as {\it casual
connection}.

The analogy with standard Quantum Mechanics is
straightforward. By considering the case in which the individual
system consists of a large number of identical subsystems, one can
derive from the above interpretation, the usual probabilistic
interpretation of quantum mechanics for the subsystems
\cite{hartle1,halliwell}.

What we proposed  is a criterion by which the Hartle
point of view can be recovered without arbitrariness. If a Noether
symmetry (or more than one) is present for a given minisuperspace
model, then strongly peaked (oscillatory) subsets of the wave
function of the universe are found. Vice-versa, oscillatory parts
of the wave function can be always connected to conserved momenta
and then to Noether symmetries.

From a general point of view, this is the same philosophy of many
branches of physics: finding symmetries allows to solve dynamics,
gives the main features of systems and simplify the interpretation
of results.

We have worked out this approach for minisuperspace models coming from Extended Theories of Gravity showing 
that identification of suitable Noether symmetries allows to completely solve the dynamical system.
Furthermore, we have seen that if a Noether symmetry is present, it is preserved by the
conformal transformation which connects Jordan and Einstein frames.
In this sense, conformally equivalent classes of minisuperspaces can be selected.

\section*{Acknowledgements}
S.C.  wants to acknowledge G. Lambiase and C. Stornaiolo for useful discussions and comments on  topics related with this paper. M.D.L. has been partially supported by  INFN-MEC bilateral agreement and {\it iniziativa specifica} NA12-INFN.  The work by S.D.O. has been supported in part by MICINN (Spain) project FIS2010-15640,  by AGAUR 2009SGR-994 project.

\end{document}